\documentclass[draftcls, onecolumn, a4paper]{IEEEtran}
\usepackage{amsmath,amssymb,amsthm,mathrsfs,amsfonts,dsfont,stmaryrd}
\usepackage{mathtools}
\usepackage{color}
\usepackage{graphicx} 
\usepackage{subfig} 
\allowdisplaybreaks[4] 

\newtheorem{theorem}{\textbf{Theorem}}

\newtheorem{proposition}{\textbf{Proposition}}

\newtheorem{definition}{\textbf{Definition}}

\newtheorem{claim}{\textbf{Claim}}
\newtheorem{remark}{\textbf{Remark}}
\newcommand{\mc}{\mathcal} 

\newcommand{\mkv}{-\!\!\!\!\minuso\!\!\!\!-}

\graphicspath{{figs/}}

\begin{document}
\title{Rate-Distortion Region of a Gray-Wyner Model with Side Information}
 \author{Meryem Benammar \qquad \qquad \qquad  Abdellatif Zaidi}
\maketitle

\begin{abstract}
In this work, we establish a full single-letter characterization of the rate-distortion region of an instance of the Gray-Wyner model with side information at the decoders. Specifically, in this model an encoder observes a pair of memoryless, arbitrarily correlated, sources $(S^n_1,S^n_2)$ and communicates with two receivers over an error-free rate-limited link of capacity $R_0$, as well as error-free rate-limited individual links of capacities $R_1$ to the first receiver and $R_2$ to the second receiver. Both receivers reproduce the source component $S^n_2$ losslessly; and Receiver $1$ also reproduces the source component $S^n_1$ lossily, to within some prescribed fidelity level $D_1$.  Also, Receiver $1$ and Receiver $2$ are equipped respectively with memoryless side information sequences $Y^n_1$ and $Y^n_2$. Important in this setup, the side information sequences are arbitrarily correlated among them, and with the source pair $(S^n_1,S^n_2)$; and are not assumed to exhibit any particular ordering. Furthermore, by specializing the main result to two Heegard-Berger models with successive refinement and scalable coding, we shed light on the roles of the common and private descriptions that the encoder should produce and what they should carry optimally. We develop intuitions by analyzing the developed single-letter optimal rate-distortion regions of these models, and discuss some insightful binary examples.   
\end{abstract}

\section{Introduction}

The Gray-Wyner source coding problem was originally formulated, and solved, by Gray and Wyner in \cite{Gray1974}. In their original setting, referred therein to as a "simple network", a pair of arbitrarily correlated memoryless sources $(S^n_1, S^n_2)$ is to be encoded and transmitted to two receivers that are connected to the encoder each through a common error-free rate-limited link as well as a private error-free rate-limited link. Because the channels are rate-limited, the encoder produces a compressed bit string $W_0$ of rate $R_0$ that it transmits over the common link, and two compressed bit strings, $W_1$ of rate $R_1$ and $W_2$ of rate $R_2$, that it transmits respectively over the private link to first receiver and the private link to the second receiver. The first receiver uses the bit strings $W_0$ and $W_1$ to reproduce an estimate, or approximation, $\hat{S}^n_1$ of the source component $S^n_1$ to within some prescribed average fidelity level $D_1$, for some distortion measure $d_1(\cdot,\cdot)$. Similarly, the second receiver uses the bit strings $W_0$ and $W_2$ to reproduce an estimate $\hat{S}^n_2$ of the source component $S^n_2$ to within some prescribed average fidelity level $D_2$, for some, possibly different, distortion measure $d_2(\cdot,\cdot)$. In~\cite{Gray1974}, Gray and Wyner characterized the optimal tradeoff among achievable rate triples $(R_0,R_1,R_2)$ and distortion pair $(D_1,D_2)$. 

\begin{figure}[ht!]
\centering
\includegraphics[scale=1]{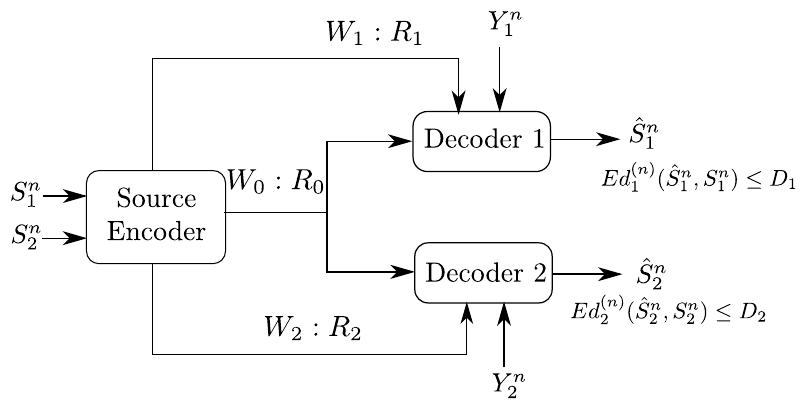}
\caption{Gray-Wyner network with side information at the receivers.}
\label{fig-Gray-Wyner-model-with-side-info}
\end{figure}

\noindent Figure~\ref{fig-Gray-Wyner-model-with-side-info} shows a generalization of Gray-Wyner's original model in which the receivers also observe, or measure, correlated memoryless side information sequences, $Y^n_1$ at Receiver $1$ and $Y^n_2$ at Receiver $2$. Some special cases of the Gray-Wyner's model with side information of Figure~\ref{fig-Gray-Wyner-model-with-side-info} have been solved (see the "Related Work" section below). However, in its most general form, i.e., when the side information sequences are arbitrarily correlated among them and with the sources, this problem has so-far eluded single-letter characterization of the optimal rate-distortion region, which is then still to be found. In fact, the problem appears somewhat hopeless as the rate-distortion function of a specific case of it, the well known Heegard-Berger problem~\cite{HB85}, which is obtained by setting $R_1=R_2=0$ in Figure~\ref{fig-Gray-Wyner-model-with-side-info}, has itself eluded the information theory for, now, more than three decades. 
 
\begin{figure}[ht!]
\centering
\includegraphics[scale=1]{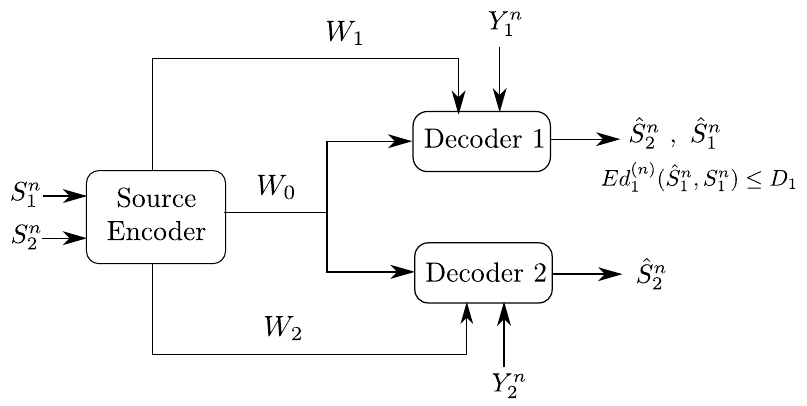}
\caption{Gray-Wyner model with side information at both receivers and degraded reconstruction sets}
\label{fig-Gray-Wyner-model-with-side-info-degraded-reconstructions}
\end{figure}

In this paper, we study an instance of the Gray-Wyner's model with side information of Figure~\ref{fig-Gray-Wyner-model-with-side-info} in which both receivers want to reproduce the source component $S^n_2$ losslessly; and Receiver $1$ also wants to reproduce the source component $S^n_1$ lossily, to within some prescribed fidelity level $D_1$. The model is shown in Figure~\ref{fig-Gray-Wyner-model-with-side-info-degraded-reconstructions}; and we refer to it as ``Gray-Wyner model with side information and degraded reconstruction sets". It is important to note that, while the reconstruction sets are assumed to be degraded, no specific ordering is imposed on the side information sequences, which then can be arbitrarily correlated among them and with the sources $(S^n_1,S^n_2)$. 

\noindent Typical to in similar problems, the encoder produces a common description of the sources pair $(S^n_1,S^n_2)$ that is intended to be recovered by both receivers, as well as individual or private descriptions of $(S^n_1,S^n_2)$ that are destined to be recovered each by a distinct receiver. Because the side information sequences do \textit{not} exhibit any specific ordering, it is not clear a-priori what information these descriptions should carry optimally. Also, it is not clear how these descriptions should be transmitted using the available common and dedicated links. 

\begin{figure*}[!h] 
			\centering  
			\subfloat[HB model with successive refinement ] 
			{
				\includegraphics[width=0.49\linewidth]{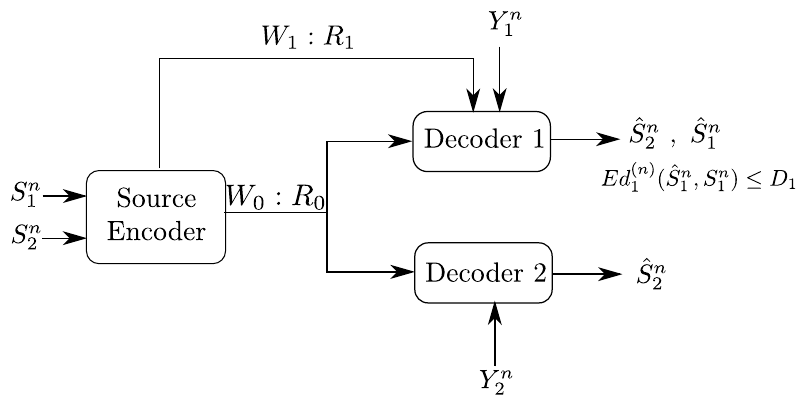}
				\label{subfig1-HB-models-successive-refinement-scalable-coding} 
			} 
		\subfloat[HB model with scalable coding] 
			{
				\includegraphics[width=0.49\linewidth]{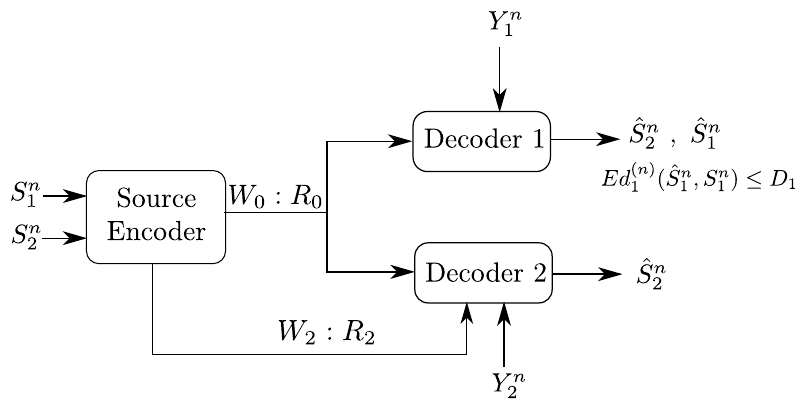} 
				\label{subfig2-HB-models-successive-refinement-scalable-coding} 
			} 
			\caption{Two classes of Heegard-Berger models} 
		\label{fig-HB-models-successive-refinement-scalable-coding} 
	\end{figure*}
	
Instrumental in the investigation of the general model of Figure~\ref{fig-Gray-Wyner-model-with-side-info-degraded-reconstructions}, and helpful for the reader to grasp the intuitions gradually, we will also study the two important underlying Heegard-Berger special models that are shown in Figure~\ref{fig-HB-models-successive-refinement-scalable-coding}. In both models, only one of the two refinement individual links has non-zero rate. In the model of Figure~\ref{subfig1-HB-models-successive-refinement-scalable-coding}, the receiver that accesses the additional rate-limited link (i.e., Receiver $1$) is also required to reproduce a lossy estimate of the source component $S^n_1$, in addition to the source component $S^n_2$ which is to be reproduced losslessly by both receivers. We will refer to this model as a ``Heegard-Berger problem with successive refinement", in reference to that if a user gets only a coarse description of the sources it should only reproduce the source component $S^n_2$ (losslessly); and if, at a later stage, a description of the source component $S^n_1$ as well is needed, the encoder sends a secondary string of compressed bits at rate $R_1$ to that user. Reminiscent of successive refinement source coding, which is a special case of the more general multiple description coding, this model may be appropriate to model applications in which descriptions of only some components (e.g., $S^n_2$) of the source suffices at the first use of the data; and descriptions of the remaining components (e.g., $S^n_1$) are needed only at a later stage. 

\noindent The model of Figure~\ref{subfig2-HB-models-successive-refinement-scalable-coding} has the individual rate-limited link connected to the receiver that is required to reproduce only the source component $S^n_2$. From an application viewpoint, the reader may find it appropriate to think about the second receiver here as having a ``lower quality" side information; and the private link is introduced precisely so that the user with the better side information and/or capability be not constrained by the communication to the user with bad and/or less good side information or capability.  We will refer to this model as a ``Heegard-Berger problem with scalable coding", reusing a term that was coined in \cite{TianDiggaviScalable} for a similar scenario, and in reference to that a user may have such strong capability (e.g., thanks to a strong side information) that only a minimal amount of information from the encoder suffices; or, conversely, barely any capability (e.g., due to weak side information) that the encoder should provide almost every information to it in order to satisfy a fidelity criterion. However, unlike~\cite{TianDiggaviScalable}, we do no make any assumption on the ordering of the side information sequences in our analysis; and the results that will follow for this model hold in general as we already mentioned.

\subsection{Main Contributions}

The main result of this paper is a single-letter characterization of the optimal rate-distortion region of the Gray-Wyner model with side information and degraded reconstruction sets of Figure~\ref{fig-Gray-Wyner-model-with-side-info-degraded-reconstructions}. To this end, in particular, we derive a converse proof that is tailored specifically for the model with degraded reconstruction sets that we study here. For the proof of the direct part, we develop a coding scheme that is very similar to one developed in the context of coding for broadcast channels with feedback in~\cite{Shayevitz2013}, but with an appropriate choice of the variables which we specify here. The specification of the main result to the Heegard-Berger models with successive refinement and scalable coding of Figure~\ref{fig-HB-models-successive-refinement-scalable-coding} sheds important light on the roles of the common and private descriptions and what they should carry optimally. We develop intuitions by analysing the established single-letter optimal rate-distortion regions of these models, and discuss some insightful binary examples.
 
\subsection{Related Work} 

In~\cite{Shayevitz2013}, Shayevitz and Wigger study a two-receiver discrete memoryless broadcast channel with feedback. They develop an efficient coding scheme which treats the feedback signal as a source that has to be conveyed lossily to the receivers in order to refine their messages' estimates, through a block Markov coding scheme. In doing so, the users' channel outputs are regarded as side information sequences; and so the scheme clearly connects with the Gray-Wyner model with side information of Figure~\ref{fig-Gray-Wyner-model-with-side-info} - as is also clearly explicit in ~\cite{Shayevitz2013}. The Gray-Wyner model with side information for which Shayevitz and Wigger's develop a (source) coding scheme, as part of their study of the broadcast channel with feedback, assumes general, possibly distinct, distortion measures at the receivers (i.e., not necessarily nested) and side information sequences that are arbitrarily correlated among them and with the source. However, its optimality is still to be shown (if it were to hold). As we already mentioned, in this paper we show that when specialized to the model with degraded reconstruction sets of Figure~\ref{fig-Gray-Wyner-model-with-side-info-degraded-reconstructions} that we study here, Shayevitz and Wigger's coding scheme for the Gray-Wyner model with side information of~\cite{Shayevitz2013} yields a rate-distortion region that meets the converse result that we here establish; and so is optimal.   

The Gray-Wyner model with side information generalizes another long standing open source coding problem, the famous Heegard-Berger problem~\cite{HeegardBerger}. Full single-letter characterization of the optimal rate-distortion function of the Heegard-Berger problem is known only in few specific cases, the most important of which are the cases of i) stochastically degraded side information sequences \cite{HeegardBerger} (see also~\cite{K94}), ii) Sgarro's result~\cite{SgarroLossless} on the corresponding lossless problem, iii) Gaussian sources with quadratic distortion measure~\cite{TianDiggaviDegraded, TianDiggaviScalable}, iv) some instances of conditionally less-noisy side information sequences~\cite{TimoLessNoisy} and v) the recently solved HB model with general side information sequences and degraded reconstruction sets~\cite{BZ15}, i.e., the model of Figure~\ref{fig-Gray-Wyner-model-with-side-info-degraded-reconstructions} with $R_1=R_2=0$ --- in the lossless case, a few other optimal results were shown, such as for the so-called complementary delivery \cite{TimoComplementary}. A lower bound for general instances of the rate distortion problem with side information at multiple decoders, that is inspired by a linear-programming lower bound for index coding, has been developed recently by Unal and Wagner in~\cite{UW16}.

\noindent Successive refinement of information was investigated by Equitz \textit{et al.} in \cite{EquitzCover91} wherein the description of the source is successively refined to a collection of receivers which are required to reconstruct the source with increasing quality levels. Extensions of successive refinement to cases in which the receivers observe some side information sequences was first investigated by Steinberg \textit{et al.} in \cite{SteinbergMerhav04} who establish the optimal rate-distortion region under the assumption that the receiver that observes the refinement link, say receiver $1$, observes also a \emph{better} side information sequence than the opposite user, i.e. the following Markov chain $S \mkv Y_1 \mkv Y_2$ holds. Tian \textit{et al.} give in \cite{TianDiggaviDegraded} an equivalent formulation of the result of \cite{SteinbergMerhav04} and extend it to the N-stage successive refinement setting. In \cite{TianDiggaviScalable}, Tian \textit{et al.} investigate another setting, for which they coined the term ``side information scalabale coding",  in which it is rather the receiver that accesses the refinement link, say receiver $2$, which observes the \emph{less good} side information sequence, i.e. $S \mkv Y_1 \mkv Y_2$. Balancing refinement quality and side information asymmetry for such a side-information scalable source coding problem allows authors in \cite{TianDiggaviScalable} to derive the rate-distortion region in the degraded side information case. The previous results on successive refinement in the presence of side information, which were generalized by Timo et al. in \cite{TimoManyUsers}, all assume, however, a specific structure in the side information sequences.

\subsection{Outline and Notation}
 An outline of the remainder of this paper is as follows. Section II describes formally the Gray-Wyner model with side information and degraded reconstruction sets of Figure~\ref{fig-Gray-Wyner-model-with-side-info-degraded-reconstructions} that we study in this paper. This section also contains some formal definitions for the Heegard-Berger type models of Figure~\ref{fig-HB-models-successive-refinement-scalable-coding}. Section III contains the main result of this paper, a full single-letter characterization of the rate-distortion region of the model of Figure~\ref{fig-Gray-Wyner-model-with-side-info-degraded-reconstructions}, together with some useful discussions and connections. A formal proof of the direct and converse parts of this result appear in Section VI. In Section IV and Section V, we specialize the result respectively to the Heegard-Berger model with successive refinement of Figure~\ref{subfig1-HB-models-successive-refinement-scalable-coding} and the Heegard-Berger model with scalable coding of Figure~\ref{subfig2-HB-models-successive-refinement-scalable-coding}. These sections also contains insightful discussions and binary examples.
  
Throughout the paper we use the following notations. The term p.m.f. stands for probability mass function. Upper case letters are used to denote random variables, e.g., $X$; lower case letters are used to denote realizations of random variables, e.g., $x$; and calligraphic letters designate alphabets, i.e., $\mc X$. Vectors of length $n$ are denoted by $X^n = (X_1, \dots, X_n)$, and $X_i^j$ is used to denote the sequence $(X_i, \dots , X_j)$. The probability distribution of a random variable $X$ is denoted by $P_X(x)\triangleq \mathds{P} (X=x)$. Sometimes, for convenience, we write it as $P_X$. We use the notation $\mathbb{E}_{X}[\cdot]$ to denote the expectation of random variable $X$. A probability distribution of a random variable $Y$ given $X$ is denoted by $P_{Y|X}$. The set of probability distributions defined on an alphabet $\mc X$ is denoted by $\mc P(\mc X)$. The cardinality of a set $\mc X$ is denoted by $\| \mc X \|$. For random variables $X$, $Y$ and $Z$, the notation $X \mkv Y \mkv Z$ indicates that $X$, $Y$ and $Z$, in this order, form a Markov Chain, i.e., $P_{XYZ}(x,y,z)= P_Y(y) P_{X|Y}(x|y)P_{Z|Y}(z|y)$. The set $\mathcal{T}^{(n)}_{[X]}$ denotes the set of sequences strongly typical with respect to the probability distribution $P_X$ and the set $\mathcal{T}^{(n)}_{[X|y^n]}$ denotes the set of sequences $x^n$ jointly typical with $y^n$ with respect to the joint p.m.f. $P_{XY}$. Throughout this paper, we use $h_2(\alpha)$ to denote the entropy of a Bernoulli\:$(\alpha)$ source, i.e., $h_2(\alpha) = - \alpha \log(\alpha) - (1-\alpha)\log(1-\alpha)$. Also, the indicator function is denoted by $\mathds{1}(\cdot)$. For real-valued scalars $a$ and $b$, with $a \leq b$, the notation $[a,b]$ means the set of real numbers that are larger or equal than $a$ and smaller or equal $b$. For integers $i \leq j$, $[i:j]$ denotes the set of integers comprised between $i$ and $j$, i.e., $[i:j] =\{i,i+1,\hdots,j\}$. Finally, throughout the paper, logarithms are taken to base $2$.

\section{Problem Setup and Formal Definitions}
 
Consider the Gray-Wyner source coding model with side information and degraded reconstruction sets shown in Figure~\ref{fig-Gray-Wyner-model-with-side-info-degraded-reconstructions}. Let $(\mathcal{S}_1 \times \mathcal{S}_2  \times \mathcal{Y}_1 \times \mathcal{Y}_2 , P_{S_1, S_2, Y_1, Y_2} )  $ be a discrete memoryless vector source with generic variables $S_1$, $S_2$, $Y_1$ and $Y_2$. Also, let $\hat{\mathcal{S}}_1$  be a reconstruction alphabet and, $d_1$ a distortion measure defined as: 
\begin{equation}
 \begin{array}{rcl}
 d_1 \ : \  \mathcal{S}_1 \times \hat{\mathcal{S}}_1 &\rightarrow&  \mathbb{R}_{+}\\ 
 (s_i, \hat{s}_i) &\rightarrow& d_1(s_i, \hat{s}_i) \ . 
 \end{array}
\label{definition-distortion}
\end{equation}

\begin{definition}
An $(n,M_{0,n},M_{1,n},M_{2,n},D_1)$ code for the Gray-Wyner source coding model with side information and degraded reconstruction sets of Figure~\ref{fig-Gray-Wyner-model-with-side-info-degraded-reconstructions} consists of:\\
- Three sets of messages $\mathcal{W}_0 \triangleq [1: M_{0,n}]$, $\mathcal{W}_1 \triangleq [1: M_{1,n}]$, and $\mathcal{W}_2 \triangleq [1: M_{2,n}]$. \\
- Three encoding functions, $f_0 $, $f_1$ and $f_2$ defined, for $j\in \{0,1,2\}$ as
\begin{equation}
 \begin{array}{rcl}
 f_j \ : \  \mathcal{S}^n_1 \times \mathcal{S}_2^n  &\rightarrow&  \mathcal{W}_j \\ 
 (S_1^n, S_2^n) &\rightarrow&  W_j = f_j ( S_1^n, S_2^n) \ . 
 \end{array}
\label{definition-encoding-function}
\end{equation}

- Two decoding functions $g_1$ and $g_2$, one at each user: 
\vspace{-1mm}
\begin{equation}
 \begin{array}{rcl}
 g_1 \ : \ \mathcal{W}_0 \times \mathcal{W}_1 \times \mathcal{Y}^n_1   &\rightarrow&   \hat{\mathcal{S}}_2^n \times \hat{\mathcal{S}}_1^n \\ 
 (W_0, W_1,  Y_1^n) &\rightarrow&  (\hat{S}_{2,1}^n, \hat{S}_1^n) = g_1 (W_0, W_1, Y_1^n ) \ , 
 \end{array}
\label{definition-decoding-function-decoder1}
\end{equation}
and 
\begin{equation}
 \begin{array}{rcl}
 g_2 \ : \ \mathcal{W}_0 \times \mathcal{W}_2 \times \mathcal{Y}^n_2   &\rightarrow&   \hat{\mathcal{S}}_2^n \\ 
 (W_0,W_2, Y_2^n) &\rightarrow& \hat{S}_{2,2}^n = g_2 (W_0, W_2, Y_2^n ) \ . 
 \end{array}
\label{definition-decoding-function-decoder2}
\end{equation} 
The expected distortion of this code is given by
\vspace{-2mm}
\begin{equation}
    \mathbb{E}\left(d^{(n)}_1(S_1^n , \hat{S}_1^n)\right) \triangleq  \mathbb{E} \dfrac{1}{n} \sum_{i=1}^n d_1( S_{1,i}, \hat{S}_{1,i})\ .  
\end{equation}  
The probability of error is defined as
\begin{equation}
 P^{(n)}_{e} \triangleq \mathds{P} \bigl(  \hat{S}_{2,1}^n \neq  S_2^n \ \text{or} \ \hat{S}_{2,2}^n \neq S_2^n\bigr)  \ .
\end{equation} 
\qed
\end{definition} 

\begin{definition}
A rate triple $(R_0, R_1, R_2)$ is said to be $D_1$-achievable for the Gray-Wyner source coding model with side information and degraded reconstruction sets of Figure~\ref{fig-Gray-Wyner-model-with-side-info-degraded-reconstructions} if there exists a sequence of $(n, M_{0,n},M_{1,n},M_{2,n}, D_1)$ codes such that: 
\begin{eqnarray}
\limsup_{n \rightarrow \infty } P^{(n)}_e &=& 0   \ , \\
\limsup_{n \rightarrow \infty } \mathbb{E}\left(d^{(n)}_1(S_1^n , \hat{S}_1^n)\right)&\leq& D_1  \ , \\
\liminf_{n \rightarrow \infty } \dfrac{1}{n}\log_2(M_{j,n})  &\leq&  R_j \   \text{for} \quad  j\in \{0,1,2\}
\end{eqnarray}
The rate-distortion region $ \mathcal{RD}$ of this problem is defined as the union of all rate-distortion quadruples $(R_0, R_1, R_2, D_1)$ such that $(R_0,R_1, R_2)$ is $D_1$-achievable, i.e, 
\begin{equation}
 \mathcal{RD} \triangleq  \cup \bigl\{ (R_0,R_1,R_2,D_1) : (R_0,R_1,R_2) \text{ is $D_1$-achievable} \bigr\} \ . 
\end{equation} \qed 
\end{definition}

As we already mentioned, we shall also study the special case Heegard-Berger type models shown in Figure~\ref{fig-HB-models-successive-refinement-scalable-coding}. The formal definitions for these models are similar to the above, and we omit them here for brevity.  

\section{Gray-Wyner Model with Side Information and Degraded Reconstruction Sets}

We establish a single-letter characterization of the optimal rate-distortion region $\mathcal{RD}$ of the Gray-Wyner model with side information and degraded reconstructions sets shown in Figure~\ref{fig-Gray-Wyner-model-with-side-info-degraded-reconstructions}. The following theorem states the result.
 
\begin{theorem}\label{th-Gray-Wyner-side-information}
The rate-distortion region $\mathcal{RD}$ of the Gray-Wyner problem with side information and degraded reconstruction set of Figure~\ref{fig-Gray-Wyner-model-with-side-info-degraded-reconstructions} is given by the sets of all rate-distortion quadruples $(R_0,R_1,R_2,D_1)$ satisfying: 
\begin{IEEEeqnarray}{rCl}
\IEEEyesnumber\IEEEyessubnumber* 
R_0 + R_1 &\geq& H(S_2|Y_1) + I(U_0 U_1; S_1 | S_2 Y_1) \\
R_0 + R_2 &\geq& H(S_2|Y_2) + I(U_0 ; S_1 | S_2 Y_2) \\
R_0 + R_1 + R_2 &\geq& H(S_2|Y_2) + I(U_0 ; S_1 | S_2 Y_2) + I(U_1; S_1 |U_0 S_2 Y_1) 
\end{IEEEeqnarray}
for some product pmf $P_{U_0 U_1 S_1 S_2 Y_1 Y_2}$, such that: 

1) the following Markov chain is valid: 
\begin{equation}
  (Y_1,Y_2) \mkv (S_1, S_2) \mkv (U_0, U_1)
 \end{equation} 

2) and there exists a function $\phi : \mathcal{Y}_1 \times \mathcal{U}_0\times \mathcal{U}_1\times \mathcal{S}_2 \rightarrow \hat{\mathcal{S}}_1$ such that:
\begin{equation}
 \mathds{E} d_1(S_1, \hat{S}_1) \leq D_1 \ . 
\end{equation}
\end{theorem} 

\noindent \textbf{Proof:} An outline of the proof of achievability of Theorem~\ref{th-Gray-Wyner-side-information} will follow. The detailed scheme and its analysis, as well as the proof of the converse part, appear in Section VI.

\noindent \textbf{Outline of Proof of Achievability:} The encoder produces a common description of $(S^n_1,S^n_2)$ that is intended to be recovered by both receivers, and an individual description that is intended to be recovered only by Receiver $1$. The common description is $V^n_0=(U^n_0,S^n_2)$; and is designed so as  to involve all of $S^n_2$, which both receivers are required to reproduce lossessly, but also all or part of  $S^n_1$, depending on the desired fidelity level $D_1$. Since we make no assumption on the side information sequences, this is meant to account for possibly unbalanced side information pairs $(Y^n_1,Y^n_2)$, in a manner that is similar to~\cite{BenammarZaidi15, BZ15}. The message that carries the common description is obtained at the encoder through the technique of double-binning developed by Tian and Diggavi in~\cite{TianDiggaviScalable}, and then used by Shayevitz and Wigger~\cite[Theorem 2]{Shayevitz2013} for a Gray-Wyner model with side information. In particular, similar to the coding scheme of ~\cite[Theorem 2]{Shayevitz2013}, the double-binning is performed in two ways, one that is relevant for Receiver $1$ and one that is relevant for Receiver $2$. Specifically, the codebook of the common description is composed of codewords $v^n_0$ that are drawn randomly and independently according to the product law of $P_{V_0}$; and is partitioned uniformly into $2^{n\tilde{R}_{0,0}}$ superbins, indexed with $\tilde{w}_{0,0} \in [1:2^{n\tilde{R}_{0,0}}]$. The codewords of each superbin of this codebook are partitioned in two distinct ways. In the first partition, they are assigned randomly and independently to $2^{n\tilde{R}_{0,1}}$ subbins indexed with $\tilde{w}_{0,1} \in [1:2^{n\tilde{R}_{0,1}}]$, according to a uniform pmf over $[1:2^{n\tilde{R}_{0,1}}]$. Similarly, in the second partition, they are assigned randomly and independently to $2^{n\tilde{R}_{0,2}}$ subbins indexed with $\tilde{w}_{0,2} \in [1:2^{n\tilde{R}_{0,2}}]$, according to a uniform pmf over $[1:2^{n\tilde{R}_{0,2}}]$. The codebook of the individual description is composed of codewords $u^n_1$ that are drawn randomly and independently according to the product law of $P_{U_1|V_0}$. This codebook is partitioned similarly uniformly into $2^{n\tilde{R}_{1,0}}$ superbins indexed with $\tilde{w}_{1,0} \in [1:2^{n\tilde{R}_{1,0}}]$, each containing $2^{n\tilde{R}_{1,1}}$ subbins indexed with $\tilde{w}_{1,1} \in [1:2^{n\tilde{R}_{1,1}}]$ codewords $u^n_1$. 

\noindent Upon observing a typical pair $(S^n_1,S^n_2)=(s^n_1,s^n_2)$, the encoder finds a pair of codewords $(v^n_0,u^n_1)$ that is jointly typical with $(s^n_1,s^n_2)$. Let $\tilde{w}_{0,0}$, $\tilde{w}_{0,1}$ and $\tilde{w}_{0,2}$ denote respectively the indices of the superbin, subbin of the first partition and  subbin of the second partition of the codebook of the common description, in which lies the found $v^n_0$. Similarly, let $\tilde{w}_{1,0}$ and $\tilde{w}_{1,1}$ denote respectively the indices of the superbin and subbin of the codebook of the individual description in which lies the found $u^n_1$. The encoder sets the common message $W_0$ as $W_0=(\tilde{w}_{0,0}, \tilde{w}_{1,0})$  and sends it over the error-free rate-limited common link of capacity $R_0$. Also, it sets the individual message $W_1$ as $W_1=(\tilde{w}_{0,1}, \tilde{w}_{1,1})$ and sends it the error-free rate-limited link to Receiver $1$ of capacity $R_1$; and the individual message $W_2$ as $W_2=\tilde{w}_{0,2}$ and sends it the error-free rate-limited link to Receiver $2$ of capacity $R_2$. For the decoding, Receiver $2$ utilizes the second partition of the codebook of the common description; and looks in the subbin of index $\tilde{w}_{0,2}$ of the superbin of index $\tilde{w}_{0,0}$ for a unique $v^n_0$ that is jointly typical with its side information $y^n_2$. Receiver $1$ decodes $v^n_0$ similarly, utilizing the first partition of the codebook of the common description and its side information $y^n_1$. It also utilizes the codebook of the individual description; and looks in the subbin of index $\tilde{w}_{1,1}$ of the superbin of index $\tilde{w}_{1,1}$ for a unique $u^n_1$ that is jointly typical with the pair $(y^n_1,v^n_0)$. In the formal proof in Section~IV, we argue that with an appropropriate choice of the communication rates $\tilde{R}_{0,0}$, $\tilde{R}_{0,1}$, $\tilde{R}_{0,2}$, $\tilde{R}_{1,0}$ and  $\tilde{R}_{1,1}$, as well as the sizes of the subbins, this scheme achieves the rate-distortion region of Theorem~\ref{th-Gray-Wyner-side-information}. \qed

\noindent A few remarks that connect Theorem~\ref{th-Gray-Wyner-side-information} to known results on related models are in order.

\begin{remark}
In the special case in which $R_1 = R_2 = 0$, the Gray-Wyner model with side information and degraded reconstruction sets of Figure~\ref{fig-Gray-Wyner-model-with-side-info-degraded-reconstructions} reduces to a Heegard-Berger problem with arbitrary side information sequences and degraded reconstruction sets, a model that was studied, and solved, recently in the authors' own recent work \cite{BZ15}. Theorem~\ref{th-Gray-Wyner-side-information} can then be seen as a generalization of~\cite[Theorem1]{BZ15} to the case in which the encoder is connected to the receivers also through error-free rate-limited private links of capacity $R_1$ and $R_2$ respectively. \qed
\end{remark} 

\begin{remark}
In~\cite{Timo08}, Timo \textit{et al.} study the Gray-Wyner source coding model with side information of Figure~\ref{fig-Gray-Wyner-model-with-side-info}. They establish the rate-region of this model in the specific case in which the side information sequence $Y^n_2$ is a degraded version of $Y^n_1$, i.e., $(S_1,S_2) \mkv Y_1 \mkv Y_2$ is a Markov chain, and both receivers reproduce the component $S^n_2$ and Receiver $1$ also reproduces the component $S^n_1$, all in a lossless manner. The result of Theorem~\ref{th-Gray-Wyner-side-information} generalizes that of~\cite[Theorem 5]{Timo08} to the case of side information sequences that are arbitrarily correlated among them and with the source pair $(S_1,S_2)$ and lossy reconstruction of $S_1$. In ~\cite{Timo08}, Timo \textit{et al.} also investigate, and solve, a few other special cases of the model, such as those of single source $S_1=S_2$~\cite[Theorem 4]{Timo08} and complementary delivery $(Y_1, Y_2) = (S_2, S_1)$~\cite[Theorem 6]{Timo08}. The results of~\cite[Theorem 4]{Timo08} and ~\cite[Theorem 6]{Timo08} can be recovered from Theorem~\ref{th-Gray-Wyner-side-information} as special cases of it. Theorem~\ref{th-Gray-Wyner-side-information} also generalizes~\cite[Theorem 6]{Timo08} to the case of lossy reproduction of the component $S^n_1$. \qed
\end{remark} 

\section{The Heegard-Berger Problem with Successive Refinement}

An important special case of the Gray-Wyner source coding model with side information and degraded reconstruction sets of Figure~\ref{fig-Gray-Wyner-model-with-side-info-degraded-reconstructions} is the case in which $R_2=0$. The resulting model, which is of successive-refinement type for a Heegard-Berger problem as we already mentioned, is shown in Figure~\ref{subfig1-HB-models-successive-refinement-scalable-coding}. In this section, we specialize the result of the previous section to this setting and discuss a binary example. It is hoped that the below analysis and discussions shed more light on the roles of the common and individual descriptions that are produced by the encoder in this case, as well as on the way they are transmitted optimally over the available links.

\subsection{Rate-Distortion Region} 

The following theorem states the optimal rate-distortion region of the Heegard-Berger problem with successive refinement of Figure~\ref{subfig1-HB-models-successive-refinement-scalable-coding}.
 
\begin{theorem}\label{th-successive-refinement}
The rate-distortion region of the Heegard-Berger problem with successive refinement of Figure~\ref{subfig1-HB-models-successive-refinement-scalable-coding} is given by the set of rate-distortion triples $(R_0, R_1, D_1)$ satisfying: 
\begin{IEEEeqnarray}{rCl}
\IEEEyesnumber\IEEEyessubnumber* 
\label{first-rate-constraint-th-successive-refinement}
R_0 &\geq& H(S_2|Y_2) + I(U_0; S_1 | S_2 Y_2) \\
\label{second-rate-constraint-th-successive-refinement}
R_0 + R_1 &\geq& H(S_2 |Y_1 ) + I(U_0 U_1; S_1 | S_2 Y_1)\\
\label{third-rate-constraint-th-successive-refinement}
R_0 + R_1 &\geq& H(S_2 |Y_2 ) + I(U_0 ; S_1 | S_2 Y_2) + I(U_1; S_1 |U_0 S_2 Y_1) 
\end{IEEEeqnarray}
for some product pmf $P_{U_0 U_1 S_1 S_2 Y_1 Y_2}$, such that: 

1) the following Markov chain is valid: 
\begin{equation}
  (U_0,U_1) \mkv (S_1, S_2) \mkv (Y_1, Y_2)
	\label{markov-chain-condition-th-successive-refinement}
 \end{equation} 

2) and there exists a function $\phi : \mathcal{Y}_1 \times \mathcal{U}_0\times \mathcal{U}_1\times \mathcal{S}_2 \rightarrow \hat{\mathcal{S}}_1$ such that:
\begin{equation}
 \mathds{E} d_1(S_1, \hat{S}_1) \leq D_1 \ . 
\label{distortion-constraint-th-successive-refinement}
\end{equation}
\end{theorem} 

\textbf{Proof:} The proof of Theorem~\ref{th-successive-refinement} follows from that of Theorem~\ref{th-Gray-Wyner-side-information} by setting $R_2=0$ therein. 

\begin{remark}
Recall the coding scheme of Theorem~\ref{th-Gray-Wyner-side-information}. If $R_2=0$, the second partition of the codebook of the common description, which is relevant for Receiver $2$, becomes \textit{degenerate} since, in this case, all the codewords $v^n_0$ of a superbin $\mc B_{00}(\tilde{w}_{0,0})$ are assigned to a single subbin. Correspondingly, the common message that the encoder sends over the common link carries only the index $\tilde{w}_{0,0}$ of the superbin $\mc B_{00}(\tilde{w}_{0,0})$ of the codebook of the common description in which lies the found typical $v^n_0=(s^n_2,u^n_0)$, in addition to the index $\tilde{w}_{1,0}$ of the subbin $\mc B_{10}(\tilde{w}_{1,0})$ of the codebook of the individual description in which lies the found typical $u^n_1$. The constraint~\eqref{first-rate-constraint-th-successive-refinement} on the common rate $R_0$ is in accordance with that Receiver $2$ utilizes only the index $\tilde{w}_{0,0}$ in the decoding. Furthermore, note that the constraints~\eqref{second-rate-constraint-th-successive-refinement} and ~\eqref{third-rate-constraint-th-successive-refinement} on the sum-rate $(R_0+R_1)$ can be combined as
\begin{equation}
R_0 + R_1 \geq \min\:\left\{ I(U_0S_2; S_1S_2|Y_1), I(U_0S_2; S_1S_2|Y_2)\right\} + I(U_1; S_1|U_0S_2Y_1) 
\end{equation} 
which readers who are familiar with Heegard-Berger type coding may recognize more easily (see, e.g., ~\cite[Theorem 2, p. 733]{HB85}).
\end{remark}

\begin{remark}\label{remark1-HB-successive-refinement}
As we already mentioned, the result of Theorem~\ref{th-successive-refinement} holds for side information sequences that are arbitrarily correlated among them and with the sources. In the specific case in which the user who gets the refinement rate-limited link also has the "better-quality" side information, in the sense that  $(S_1,S_2) \mkv Y_1 \mkv Y_2$ forms a Markov chain, the rate-distortion region of Theorem~\ref{th-successive-refinement} reduces to
the set of all rate-distortion triples $(R_0,R_1,D_1)$ that satisfy
\begin{IEEEeqnarray}{rCl}\label{scalable-reconstruction-degraded-side-info}
\IEEEyesnumber\IEEEyessubnumber* 
R_0 &\geq& H(S_2|Y_2) + I(U_0; S_1 | S_2 Y_2) \\ 
R_0 + R_1 &\geq& H(S_2 |Y_2 ) + I(U_0 ; S_1 | S_2 Y_2) + I(U_1; S_1 |U_0 S_2 Y_1)  \ . 
\end{IEEEeqnarray}
for some joint measure $P_{U_0 U_1 S_1 S_2 Y_1 Y_2}$ for which \eqref{markov-chain-condition-th-successive-refinement} and \eqref{distortion-constraint-th-successive-refinement} hold. This result can also be obtained from previous works on successive refinement for the Wyner-Ziv source coding problem by Steinberg and Merhav ~\cite[Theorem 1]{SteinbergMerhav04} and Tian and Diggavi~\cite[Theorem 1]{TianDiggaviDegraded}. The results of~\cite[Theorem 1]{SteinbergMerhav04} and~\cite[Theorem 1]{TianDiggaviDegraded} hold for possibly distinct, i.e., not necessarily nested, distortion measures at the receivers; but they require the aforementioned Markov chain condition which is pivotal for their proofs. Thus, for the considered degraded reconstruction sets setting, Theorem~\ref{th-successive-refinement} can be seen as generalizing~\cite[Theorem 1]{SteinbergMerhav04} and~\cite[Theorem 1]{TianDiggaviDegraded} to the case in which the side information sequences are arbitrarily correlated among them and with the sources $(S_1,S_2)$, i.e., do not exhibit any ordering.  \qed

\end{remark}

\begin{remark}\label{remark2-HB-successive-refinement}
In the case in which it is the user who gets only the common rate-limited link that has the "better-quality" side information, in the sense that $(S_1,S_2) \mkv Y_2 \mkv Y_1$ forms a Markov chain, the rate distortion region of Theorem~\ref{th-successive-refinement} reduces to the set of all rate-distortion triples $(R_0, R_1, D_1)$ that satisfy
\begin{IEEEeqnarray}{rCl}\label{scalable-reconstruction-scalable-side-info}
\IEEEyesnumber\IEEEyessubnumber* 
R_0 &\geq& H(S_2|Y_2) + I(U_0; S_1 | S_2 Y_2) \\
R_0 + R_1 &\geq& H(S_2 |Y_1) + I(U_0 U_1; S_1 | S_2 Y_1) 
\end{IEEEeqnarray}
for some joint measure $P_{U_0 U_1 S_1 S_2 Y_1 Y_2}$ for which \eqref{markov-chain-condition-th-successive-refinement} and \eqref{distortion-constraint-th-successive-refinement} hold. This result can also be conveyed from~\cite{TianDiggaviScalable}. Specifically, in ~\cite{TianDiggaviScalable} Tian and Diggavi study a therein referred to as "side-information scalable" source coding setup where the side informations are degraded, and the encoder produces two descriptions such that the receiver with the better-quality side information (Receiver $2$ if $(S_1,S_2) \mkv Y_2 \mkv Y_1$ is a Markov chain) uses only the first description to reconstruct its source while the receiver with the low-quality side information (Receiver $1$ if $(S_1,S_2) \mkv Y_2 \mkv Y_1$ is a Markov chain) uses the two descriptions in order to reconstruct its source. They establish inner and outer bounds on the rate-distortion region of the model, which coincide when either one of the decoders requires a lossless reconstruction or when the distortion measures are degraded and deterministic. Similar to in the previous remark, Theorem~\ref{th-successive-refinement} can be seen as generalizing the aforementioned results of~\cite{TianDiggaviScalable} to the case in which the side information sequences are arbitrarily correlated among them and with the sources $(S_1,S_2)$.\qed 
\end{remark} 

\subsection{Binary Example}
\begin{figure}[ht!]
\centering
\includegraphics[scale=1]{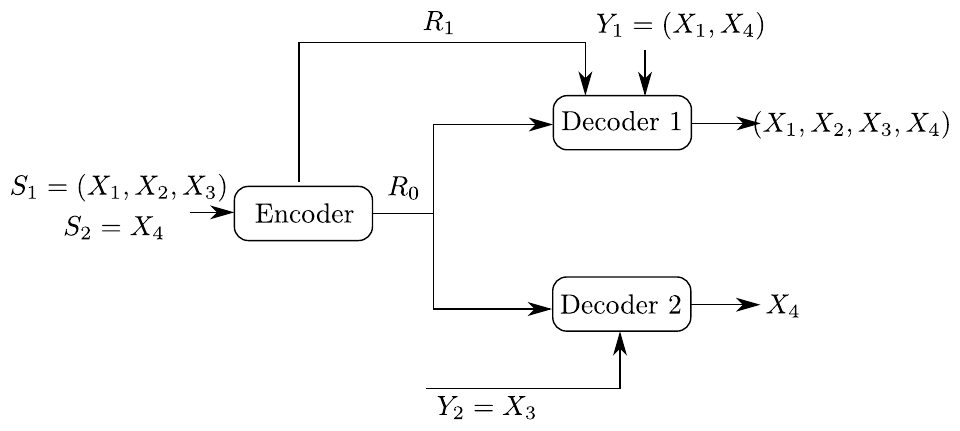}
\caption{Binary Heegard-Berger example with successive refinement}
\label{fig-binary-example-HB-with-successive-refinement}
\end{figure}

Let $X_1$, $X_2$, $X_3$ and $X_4$ be four independent $\text{Ber}(1/2)$ random variables. Let the sources be $S_1 \triangleq (X_1,X_2,X_3)$ and $S_2 \triangleq X_4$. Now, consider the Heegard-Berger model with successive refinement shown in Figure~\ref{fig-binary-example-HB-with-successive-refinement}. The first user, which gets both the common and individual links,  observes the side information $Y_1=( X_1, X_4)$ and wants to reproduce the pair $(S_1,S_2)$ losslessly. The second user gets only the common link, has side information $Y_2=(X_2,X_3)$ and wants to reproduce only the component $S_2$, losslessly.

\noindent It is easy to check that the side information at the decoders do \textit{not} exhibit any degradedness ordering, in the sense that none of the Markov chain conditions of Remark~\ref{remark1-HB-successive-refinement} and Remark~\ref{remark2-HB-successive-refinement} hold. The following claim provides the rate-region of this binary example.

\begin{claim}\label{claim-binary-example-HB-successive-refinement}
The rate region of the binary Heegard-Berger example with successive refinement of Figure~\ref{fig-binary-example-HB-with-successive-refinement} is given by the set of rate pairs $(R_0,R_1)$ that satisfy
\begin{IEEEeqnarray}{rCl}\label{ex-rate-region-successive-refinement-02}
\IEEEyesnumber\IEEEyessubnumber* 
R_0 &\geq&  1 \\
R_0 + R_1 &\geq& 2  \ .  
\end{IEEEeqnarray}
\end{claim}

\textbf{Proof:} The proof of Claim~\ref{claim-binary-example-HB-successive-refinement} follows easily by specializing, and computing, the result of Theorem~\ref{th-successive-refinement} for the example at hand, as follows. First note that for $D_1=0$ and the variables $(S_1,S_2,Y_1,Y_2)$ chosen as in the example, the inequality~\eqref{second-rate-constraint-th-successive-refinement} gives
\begin{subequations}
\begin{align}
R_0 + R_1 &\geq H(S_2 |Y_1 ) + I(U_0 U_1; S_1 | S_2 Y_1) \\
&= H(S_1S_2|Y_1)  \\
&= 2
\end{align}
\label{constraint1-sum-rate-proof-claim1}
\end{subequations}
where the first equality follows since $H(S_1|S_2Y_1U_0U_1)=0$ for all random variables $U_0$ and $U_1$ that satisfy \eqref{markov-chain-condition-th-successive-refinement} and \eqref{distortion-constraint-th-successive-refinement}, by the lossless reconstruction of the source component $S_1$ at the first receiver. Also, the inequality~\eqref{first-rate-constraint-th-successive-refinement} gives
\begin{subequations}
\begin{align}
R_0 &\geq H(S_2|Y_2) + I(U_0; S_1 | S_2 Y_2) \\
&= 2 - H(X_1| X_2 X_3 X_4 U_0) \\
&\geq 2 - H(X_1| X_2 X_3 X_4) \\
&= 1
\end{align}
\label{constraint-common-rate-proof-claim1}
\end{subequations}
and the inequality~\eqref{third-rate-constraint-th-successive-refinement} gives
\begin{subequations}
\begin{align}
R_0 + R_1 &\geq H(S_2 |Y_2 ) + I(U_0 ; S_1 | S_2 Y_2) + I(U_1; S_1 |U_0 S_2 Y_1)  \\
&= 2 + H(X_2 X_3 |X_1 X_4 U_0) - H(X_1| X_2 X_3 X_4 U_0) \\
&= 2 + \left[ H(X_2 X_3 |X_4 U_0) - H(X_1| X_4 U_0) \right]
\end{align}
\label{constraint2-sum-rate-proof-claim1}
\end{subequations}
for some $U_0$ that satisfies \eqref{markov-chain-condition-th-successive-refinement} and \eqref{distortion-constraint-th-successive-refinement}. Next, combining \eqref{constraint1-sum-rate-proof-claim1} and \eqref{constraint2-sum-rate-proof-claim1}, we get
\begin{subequations}
\begin{align}
R_0 + R_1 &\geq 2 + \max\{0, H(X_2 X_3 |X_4 U_0) - H(X_1| X_4 U_0) \} \\
 &\geq 2.
\end{align}
\label{equivalent-constraint-sum-rate-proof-claim1}
\end{subequations}
The end of the proof of the claim follows by noticing that the inequalities in \eqref{constraint-common-rate-proof-claim1} and \eqref{equivalent-constraint-sum-rate-proof-claim1} hold with equality with the choice $U_0=(X_2,X_3)$. \qed

\begin{figure}[h!]
\centering 
\includegraphics[scale=.4]{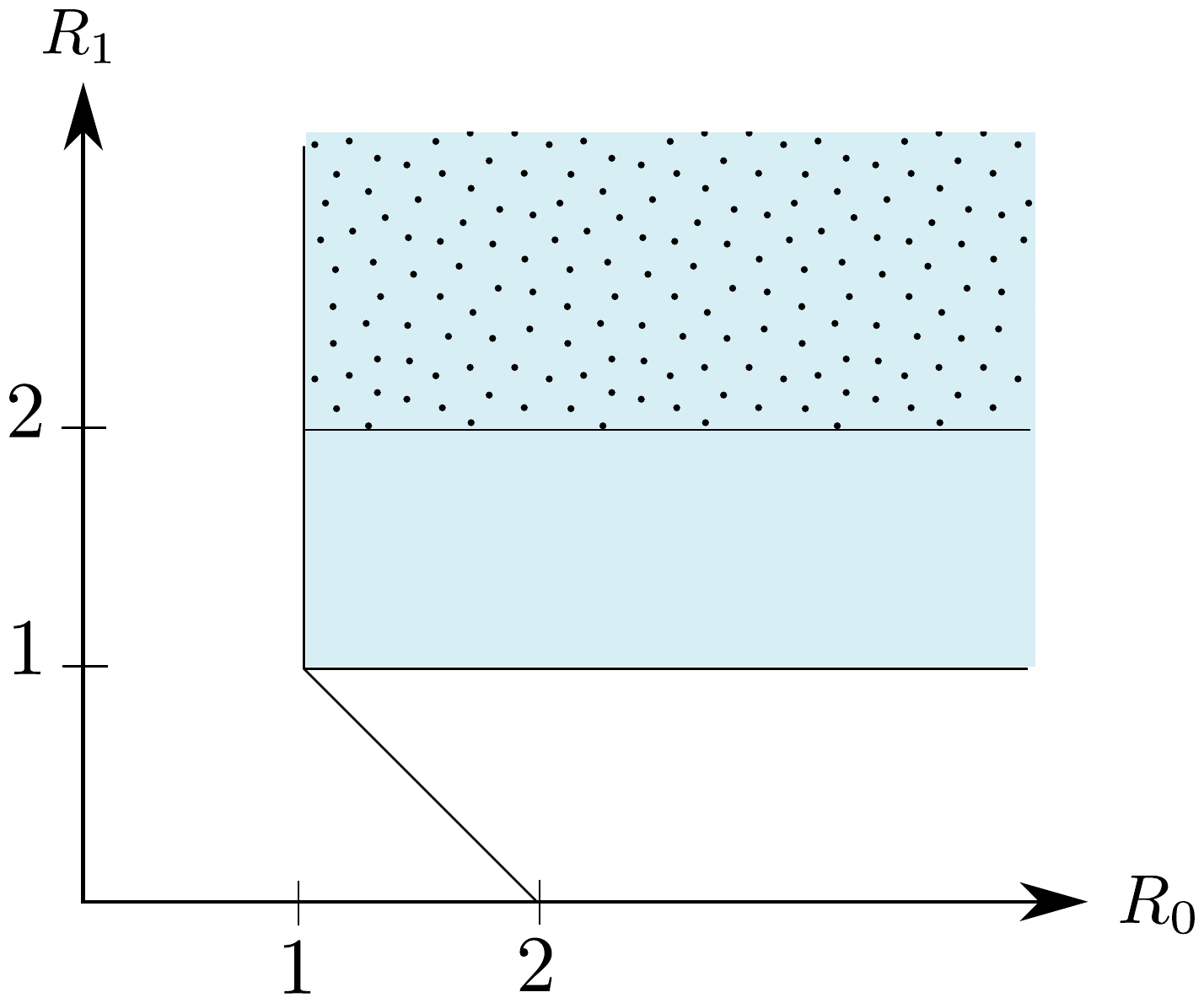}
\caption{Rate region of the binary example of Figure~\ref{fig-binary-example-HB-with-successive-refinement}.}
\label{fig-ex-successive-refinement}
\end{figure}

\noindent The rate region of Claim~\ref{claim-binary-example-HB-successive-refinement} is depicted in Figure~\ref{fig-ex-successive-refinement}. It is insightful to notice that although the second user is only interested in reproducing the component $S_2=X_4$, the optimal coding scheme that achieves this region sets the common description that is destined to be recovered by both users as one that is composed of not only $S_2$ but also some part $U_0=(X_2,X_3)$ of the source component $S_1$ (though the latter is not required by the second user). A possible intuition is that the choice $U_0=(X_2,X_3)$ is useful for user 1, who wants to reproduce $S_1=(X_1,X_2,X_3)$, and its transmission to also the second user does not cost any rate loss since this user has available side information $Y_2=(X_2,X_3)$. 

\begin{remark}
The above observation on that it is generally beneficial that the common description to be recovered by both users contains all or part of the source component $S_1$ was also made in the authors' own recent work~\cite{BZ15} (see also~\cite{BenammarZaidi15}) in the context of a Heegard-Berger problem with degraded reconstructions sets and no successive refinement, i.e., $R_1=0$. In the present model, however, the intuition is less easy to grasp, because of the presence of the refinement link. For instance, the reader may wonder whether it is equally optimal that every information that is of interest for only the first user (e.g., $(X_2,X_3)$ in the binary example of Figure~\ref{fig-binary-example-HB-with-successive-refinement}) be sent only over the dedicated private link of that user. Our analysis shows that while this is sometimes equally optimal, it is not in general; and, in fact, roughly speaking the optimal choice of the common description layer (and, so, of the auxiliary random variable $U_0$) depends on the relative strength of the pair $(R_1,Y_1)$ in comparison with $Y_2$. For example, consider again the binary example of Figure~\ref{fig-binary-example-HB-with-successive-refinement}. For this example, the lossless transmission of the source component $S_1$ to the first user requires $H(S_1|S_2Y_1)=2$ bits/sample. If the capacity of the rate-limited private link suffices to convey such information, i.e., $R_1 \geq 2$ bits/sample, then the common-description needs not contain any part of $S_1$ and the choice $U_0=\emptyset$ is equally optimal. If, however, the rate-limited private link is not good enough to convey the required information, e.g., $1 \leq R_1 < 2$ bits/sample, the encoder should transmit part of $S_1$ as part of the common description, e.g., through the choices $U_0=X_2$ or $U_0=X_3$ which can easily be shown to be equally optimal in this case. \qed
\end{remark}

\section{The Heegard-Berger Problem with Scalable Coding}

Consider now the model of Figure~\ref{subfig2-HB-models-successive-refinement-scalable-coding}. As we already mentioned, the reader may find it appropriate for the motivation to think about the side information $Y^n_2$ as being of lower quality than $Y^n_1$, in which case the refinement link that is given to the second user is in accordance with the motivation that the receiver with the better-quality side information (Receiver $1$) uses only the common description layer to decode while the receiver with the low-quality side information (Receiver $2$) needs the two descriptions to decode. The results that will follow, however, hold for general side information sequences.

\subsection{Rate-Distortion Region}
The following theorem states the rate-distortion region of the Heegard-Berger model with scalable coding of Figure~\ref{subfig2-HB-models-successive-refinement-scalable-coding}.

\begin{theorem}\label{th-scalable-reconstruction}
The rate-distortion region of the Heegard-Berger model with scalable coding of Figure~\ref{subfig2-HB-models-successive-refinement-scalable-coding} is given by the set of all rate-distortion triples $(R_0,R_2,D_1)$ that satisfy 
\begin{IEEEeqnarray}{rCl}
\IEEEyesnumber\IEEEyessubnumber* 
\label{first-rate-constraint-th-scalable-reconstruction}
R_0 &\geq& H(S_2|Y_1) + I(U_0 U_1; S_1 | S_2 Y_1) \\ 
R_0 + R_2 &\geq&  H(S_2 |Y_2) + I(U_0 ; S_1 | S_2 Y_2) + I(U_1; S_1 |U_0 S_2 Y_1) 
\label{second-rate-constraint-th-scalable-reconstruction}
\end{IEEEeqnarray}
for some product pmf $P_{U_0 U_1 S_1 S_2 Y_1 Y_2}$, such that: 

1) the following Markov chain is valid: 
\begin{equation}
  (U_0,U_1) \mkv (S_1, S_2) \mkv (Y_1, Y_2)
	\label{markov-chain-condition-th-scalable-reconstruction}
 \end{equation} 

2) and there exists a function $\phi : \mathcal{Y}_1 \times \mathcal{U}_0\times \mathcal{U}_1\times \mathcal{S}_2 \rightarrow \hat{\mathcal{S}}_1$ such that:
\begin{equation}
 \mathds{E} d_1(S_1, \hat{S}_1) \leq D_1 \ . 
\label{distortion-constraint-th-scalable-reconstruction}
\end{equation}
\end{theorem}  

\textbf{Proof:} The proof of Theorem~\ref{th-scalable-reconstruction} follows from that of Theorem~\ref{th-Gray-Wyner-side-information} by seeting $R_1=0$ therein. 

\begin{remark}\label{remark2-HB-scalable-coding}
In the case in which the source component $S_1$ as well is recovered losslessly at Receiver $1$, the rate distortion region of Theorem \ref{th-successive-refinement} reduces to the set of rate pairs $(R_0,R_1)$ that satisfy
\begin{subequations}
\begin{align} 
\label{constraint-common-rate-th-scalable-reconstruction-lossless-case}
R_0 &\geq H(S_1S_2|Y_1) \\ 
R_0 + R_2 &\geq  H(S_1S_2 |Y_2) + \left[ H(S_1 | U_0S_2 Y_1) - H(S_1 | U_0S_2 Y_2)\right] 
\label{constraint-sum-rate-th-scalable-reconstruction-lossless-case}
\end{align}
\label{rate-region-th-scalable-reconstruction-lossless-case}
\end{subequations}
for some product pmf $P_{U_0|S_1S_2}$. As we already mentioned, the result of Theorem~\ref{th-scalable-reconstruction}, and so also the rate region described by ~\eqref{rate-region-th-scalable-reconstruction-lossless-case} in the lossless case, hold for general, arbitrarily correlated, side information pairs $(Y_1,Y_2)$. The rate region given by~\eqref{rate-region-th-scalable-reconstruction-lossless-case} generalizes that of~\cite[Theorem 4, item (1)]{TianDiggaviScalable}, that is developed by Tian and Diggavi for the case in which  $(S_1,S_2) \mkv Y_1 \mkv Y_2$ forms a Markov chain, to the case of arbitrarily correlated side information pairs $(Y_1,Y_2)$. \qed
\end{remark}

\begin{remark}\label{remark1-HB-scalable-coding} 
 In the specific case in which Receiver $2$ has a better-quality side information in the sense that $(S_1,S_2) \mkv Y_2 \mkv Y_1$ forms a Markov chain, the rate distortion region of Theorem \ref{th-scalable-reconstruction} reduces to one that is described by a single rate-constraint, namely 
\begin{IEEEeqnarray}{rCl}
R_0 &\geq& H(S_2|Y_1) + I(U; S_1 | S_2 Y_1) 
\end{IEEEeqnarray}
for some conditional $P_{U|S_1S_2}$ that satisfies $\mathbb{E}[d_1(S_1,\hat{S}_1] \leq D_1$. This is in accordance with the observation that, in this case, the transmission to Receiver $1$ becomes the bottleneck, as Receiver $2$ can recover the source component $S_2$ losslessly as long as so does Receiver $1$. 
\end{remark}

\begin{remark}
In~\cite[Theorem 1]{TimoManyUsers}, Timo \text{et al.} present an achievable rate-region for the multistage successive-refinement problem with side information. Timo  \text{et al.} consider distortion measures of the form $\delta_l\: :\: \mc X {\times} \hat{\mc X}_l \rightarrow \mathbb{R}_{+}$, where $\mc X$ is the source alphabet and $\hat{\mc X}_l$ is the reconstruction at decoder $l$, $l \in \{1,\hdots,t\}$; and for this reason this result is not applicable as is to the setting of Figure~\ref{subfig2-HB-models-successive-refinement-scalable-coding}, in the case of two decoders. However, the result of~\cite[Theorem 1]{TimoManyUsers} can be extended to accomodate a distortion measure at the first decoder that is vector-valued; and the direct part of Theorem \ref{th-scalable-reconstruction} can then be obtained by applying this extension. Specifically, in the case of two decoders, i.e., $t=2$, and with $X=(S_1,S_2)$, and two distortion measures $\delta_1 : \mc S_1 \times \mc S_2 \times \hat{\mc S}_{1,1} \times \hat{\mc S}_{1,2} \rightarrow \{0,1\} \times \mathbb{R}_{+}$ and $\delta_2 : \mc S_1 \times \mc S_2 \times \hat{\mc S}_{1,2} \times \hat{\mc S}_{2,2} \rightarrow \{0,1\}$
 chosen such that 
\begin{equation}
\delta_1\Big((s_1,s_2), (\hat{s}_{1,1},\hat{s}_{2,1})\Big) = \Big(d_H(s_2, \hat{s}_{2,1}), d_1(s_1,\hat{s}_{1,1})\Big)
\end{equation}
and 
\begin{equation}
\delta_2\Big((s_1,s_2), (\hat{s}_{1,2},\hat{s}_{2,2})\Big) = d_H(s_2, \hat{s}_{2,2})
\end{equation}
where $d_H(\cdot,\cdot)$ is the Hamming distance, letting $d_1=(0,D_1)$ and $d_2=0$, a straightforward extension of ~\cite[Theorem 1]{TimoManyUsers} to this setting yields a rate-region that is described by the following rate constraints (using the notation of ~\cite[Theorem 1]{TimoManyUsers})  

\begin{subequations}
\begin{align}
 R_0 &\geq  \Phi(\mathcal{T}_0, 1) +  \Phi(\mathcal{T}_1, 1) \\
 R_0 + R_2 &\geq \Phi(\mathcal{T}_0, 2) +  \Phi(\mathcal{T}_1, 2) +  \Phi(\mathcal{T}_2, 2) 
\end{align}
\label{rate-region1-Timo-multistage-successive-refinement}
\end{subequations}
where $\mc T_{0}=\{1,2\}$, $\mc T_1=\{1\}$, $T_2=\{2\}$,  and for $j=0,1,2$ and $l \in {1,2}$ such that $\mc T_j \cap \{1,\hdots,l\} \neq \emptyset$, the function $\Phi(\mathcal{T}_j, l)$, $j=0,1,2$, is defined as 
\begin{equation}
 \Phi(\mathcal{T}_j, l)= I\left(S_1S_2 \mathcal{A}_{\mathcal{T}_j}^{\dagger}; U_{\mathcal{T}_j} |\mathcal{A}_{\mathcal{T}_j}^{\supset} \right) - \displaystyle\min_{ l^\prime \in \mathcal{T}_j \cap [1:l]} I\left(  U_{\mathcal{T}_j} ;\mathcal{A}_{\mathcal{T}_j,l^\prime}^{\ddagger} Y_{l^\prime} |\mathcal{A}_{\mathcal{T}_j}^{\supset}  \right) 
\label{function-phi-rate-region-Timo-multistage-successive-refinement}
 \end{equation}
where $\mc A=\{U_{12}, U_1, U_2\}$ and the sets $\mathcal{A}_{\mathcal{T}_j}^-$, $\mathcal{A}_{\mathcal{T}_j}^{\supset}$, $\mathcal{A}_{\mathcal{T}_j}^+$, $\mathcal{A}_{\mathcal{T}_j}^{\dagger}$,  $\mathcal{A}_{\mathcal{T}_j,1}^{\ddagger}$, $\mathcal{A}_{\mathcal{T}_j,2}^{\ddagger}$, evaluated in this case, are given in Table~\ref{table1}. It is easy to see that the region described by~\eqref{rate-region1-Timo-multistage-successive-refinement} can be written more explicitly in this case as 
\begin{subequations}
\begin{align}
 R_0 &\geq I(U_{12}; S_1S_2 | Y_1)\\
 R_0 + R_2 &\geq  \max\{I(U_{12}; S_1 S_2|Y_1),I(U_{12}; S_1 S_2|Y_2) \} +  I(U_1; S_1 S_2 | Y_1 U_{12} ) + I(U_2; S_1 S_2 | Y_2 U_{12} ) \ . 
\end{align}
\label{rate-region2-Timo-multistage-successive-refinement}
\end{subequations}
Also, setting $U_{12}=(U_0,S_2)$ and $U_2=S_2$ in \eqref{rate-region2-Timo-multistage-successive-refinement} one recovers the rate-region of Theorem~\ref{th-scalable-reconstruction}. ( Such a connection can also be stated for the result of Theorem \ref{th-successive-refinement} ). 
\end{remark}

\begin{table}
\centering
\begin{tabular}{|c|c|c|c|}
\hline 
  & $\mathcal{T}_0$ & $\mathcal{T}_1$ & $\mathcal{T}_2$ \\ 
\hline 
$\mathcal{A}_{\mathcal{T}_j}^-$ & $\emptyset$ & $\emptyset$ & $U_1$\\ 
\hline 
$\mathcal{A}_{\mathcal{T}_j}^{\supset}$ & $\emptyset$ & $U_{12}$ & $U_{12}$ \\ 
\hline 
$\mathcal{A}_{\mathcal{T}_j}^+$  & $\{U_1, U_2 \}$ & $\emptyset$ & $\emptyset$ \\ 
\hline 
$\mathcal{A}_{\mathcal{T}_j}^{\dagger}$  & $\emptyset$& $\emptyset$ & $\emptyset$ \\ 
\hline 
$\mathcal{A}_{\mathcal{T}_j,1}^{\ddagger} $ & $\emptyset$ & $\emptyset$ & $\emptyset$ \\ 
\hline 
$\mathcal{A}_{\mathcal{T}_j,2}^{\ddagger}$ & $\emptyset$ & $\emptyset$ & $\emptyset$ \\ 
\hline 
\end{tabular}
\caption{Auxiliary random variables associated with the subsets that appear in \eqref{function-phi-rate-region-Timo-multistage-successive-refinement}.} 
\label{table1} 
\end{table}

\subsection{Binary Example}

Let $X_1$, $X_2$ and $X_3$ be three independent $\text{Ber}(1/2)$ random variables. Let now the sources be such that $S_1 \triangleq (X_1,X_3)$ and $S_2 \triangleq X_2$. Now, consider the Heegard-Berger model with scalable coding shown in Figure~\ref{fig-binary-example-HB-with-successive-refinement}. The first user gets only the common rate-limited link; and it  observes the side information $Y_1=( X_1, X_2)$ and wants to reproduce the pair $(S_1,S_2)$ losslessly. The second user gets both the common and refinement rate-limited links; and it observes the side information $Y_2=X_3$ and wants to reproduce only the component $S_2$, losslessly.

\begin{figure}[ht!]
\centering
\includegraphics[scale=1]{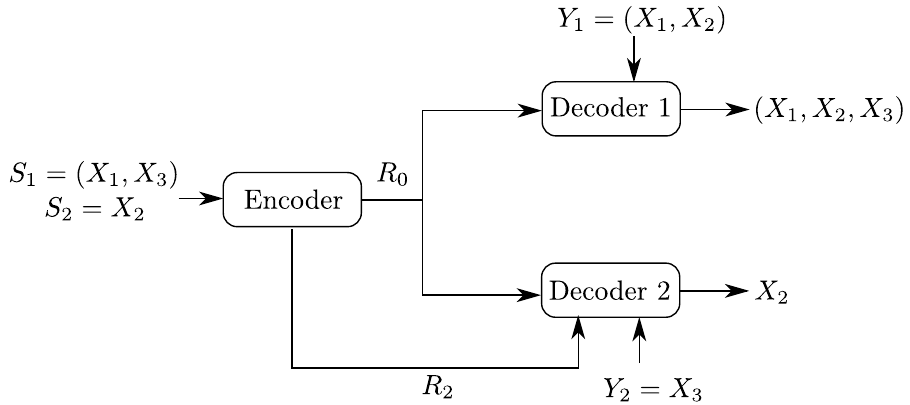}
\caption{Binary Heegard-Berger example with scalable coding}
\label{fig-binary-example-HB-with-scalable-coding}
\end{figure}

\noindent For this example as well, it is easy to see that the side information at the decoders do \textit{not} exhibit any degradedness ordering, in the sense that none of the Markov chain conditions of Remark~\ref{remark1-HB-scalable-coding} and Remark~\ref{remark2-HB-scalable-coding} hold. The following claim provides the rate-region of this binary example.

\begin{claim}\label{claim-binary-example-HB-scalable-coding}
The rate region of the binary Heegard-Berger example with scalable coding of Figure~\ref{fig-binary-example-HB-with-scalable-coding} is given by the set of all rate pairs $(R_0,R_1)$ that satisfy $R_2 \geq 0$ and $R_0 \geq 1$.
\end{claim}

\textbf{Proof:} The proof of Claim~\ref{claim-binary-example-HB-scalable-coding} follows easily by specializing, and computing, the result of Theorem~\ref{th-scalable-reconstruction} for the example at hand, as follows. First note that for $D_1=0$ and the variables $(S_1,S_2,Y_1,Y_2)$ chosen as in this example, the constraint~\eqref{first-rate-constraint-th-scalable-reconstruction} on the rate of the common link becomes
\begin{subequations}
\begin{align}
R_0  &\geq H(S_2 |Y_1 ) + I(U_0 U_1; S_1 | S_2 Y_1) \\
&= H(S_1S_2|Y_1)  \\
&= 1
\end{align}
\label{constraint-common-rate-proof-claim2}
\end{subequations}
where the first equality follows since $H(S_1|S_2Y_1U_0U_1)=0$ for all random variables $U_0$ and $U_1$ that satisfy \eqref{markov-chain-condition-th-scalable-reconstruction} and \eqref{distortion-constraint-th-scalable-reconstruction}, by the lossless reconstruction of the source component $S_1$ at the first receiver. Also, the constraint~\eqref{second-rate-constraint-th-scalable-reconstruction} on the sum of the rates on the two links becomes
\begin{subequations}
\begin{align}
R_0 + R_2 &\geq  H(S_2 |Y_2 ) + I(U_0 ; S_1 | S_2 Y_2) + I(U_1; S_1 |U_0 S_2 Y_1)  \\
& = 2 + H( X_3 |X_1 X_2 U_0) - H(X_1| X_2 X_3 U_0) \\
&=  2 + \left[H( X_3 | X_2 U_0) - H(X_1| X_2 U_0) \right] \\
&\geq 2 -  H(X_1| X_2 U_0) \\
&\geq 1
\end{align}
\label{constraint-sum-rate-proof-claim2}
\end{subequations}
where the last inequality follows since $H(X_1| X_2 U_0) \leq H(X_1)=1$ for all choices of $U_0$ that satisfy \eqref{markov-chain-condition-th-scalable-reconstruction} and \eqref{distortion-constraint-th-scalable-reconstruction}. 

\noindent The end of the proof of the claim follows by noticing that the last two inequalities on the RHS of \eqref{constraint-sum-rate-proof-claim2} hold with equality with the choice $U_0=X_3$. \qed

\begin{figure}[h!]
\centering 
\includegraphics[scale=.4]{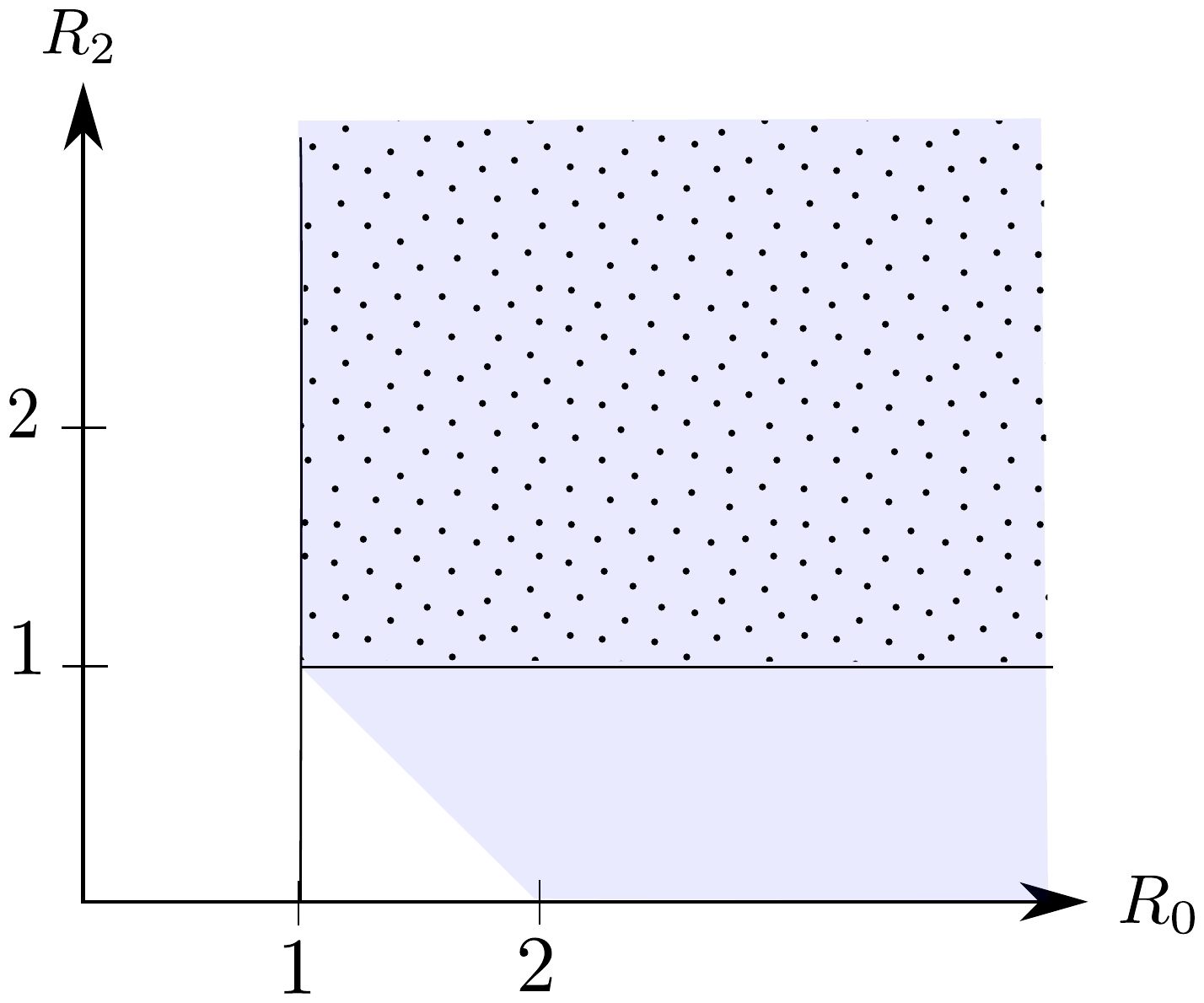}
\caption{Optimal rate region}
\label{fig-ex-scalable-coding}
\end{figure}

\noindent The optimal rate region of Claim~\ref{claim-binary-example-HB-scalable-coding} is depicted in Figure~\ref{fig-ex-scalable-coding}, as the region delimited by the lines $R_0=1$ and $R_2=0$. Note that for this example, the source component $X_2$, which is the only source component that is required by Receiver $2$, needs to be transmitted entirely on the common link so as to be recovered losslessly also by Receiver $1$. For this reason, the refinement link is not-constrained and appears to be useless for this example. The reader may notice here the sharp difference with binary Heegard-Berger example with successive refinement of Figure~\ref{fig-binary-example-HB-with-successive-refinement} for which the refinement link may sometimes be instrumental to reducing the required rate on the common link. Also, it is insightful to notice that for this example, because of the side information configuration, the choice $U_0=\emptyset$ in Theorem~\ref{th-scalable-reconstruction} is strictly suboptimal and results in the smaller region that is described by
\begin{IEEEeqnarray}{rCl}\label{suboptimal-rate-region-scalable-coding}
\IEEEyesnumber\IEEEyessubnumber* 
R_0 &\geq&  1 \\ 
R_0 + R_2 & \geq&  2.
\end{IEEEeqnarray}

\section{Proof of Theorem \ref{th-Gray-Wyner-side-information}}\label{pp-proof-th-Gray-Wyner-side-information}

\subsection{Proof of Converse Part}
Assume that a rate triple $(R_0, R_1, R_2)$ is $D_1$-achievable. Let then $W_j = f_j(S_1^n,S_2^n)$, where $j \in \{0,1,2\}$, be the encoded indices and let $\hat{S}_1^n = g_1(W_0, W_1 , Y_1^n)$ be the reconstruction sequence at the first decoder such that $\mathds{E} d^{(n)}_1(S_1^n, \hat{S}_1^n) \leq D_1$.

Using Fano's inequality, the lossless reconstruction of the source $S_2^n$ at both decoders implies that there exists a sequence $\epsilon_n \underset{n \rightarrow \infty }{\rightarrow} 0 $  such that:
\begin{IEEEeqnarray}{rCl}
 H(S_2^n | W_0 W_1 Y_1^n ) &\leq& n\epsilon_n \label{Fano-inequality-1} \\
 H(S_2^n | W_0 W_2 Y_2^n ) &\leq& n\epsilon_n \label{Fano-inequality-2}  
\end{IEEEeqnarray}

\noindent We start by showing the following sum-rate constraint, 
\begin{equation}
 R_0 + R_1+ R_2 \geq  H(S_2 |Y_2 ) + I(U_0 ; S_1 | S_2 Y_2) + I(U_1; S_1 |U_0 S_2 Y_1)  \ . 
\end{equation}

We have that
\begin{IEEEeqnarray}{rCl}
\IEEEyesnumber\IEEEyessubnumber* 
&& n (R_0 + R_1 + R_2) \IEEEnonumber\\
&\geq& H(W_0) + H(W_2) +  H(W_1) \\
&\geq& H(W_0) + H(W_2|W_0) +  H(W_1) \\
&=& H(W_0 W_2) + H(W_1) \\ 
&\geq& H(W_0 W_2| Y_2^n ) + H(W_1|W_0 S_2^n Y_1^n ) \\
&\geq& I(W_0 W_2; S_1^n S_2^n |Y_2^n) + I(W_1; S_1^n | W_0  S_2^n Y_1^n) \\ 
&=& H( S_1^n S_2^n |Y_2^n) - H(S_1^n S_2^n | W_0 W_2 Y_2^n ) + H( S_1^n | W_0 S_2^n Y_1^n) - H( S_1^n | W_0  W_1 S_2^n Y_1^n)  \\ 
&\overset{(a)}{\geq}& H( S_1^n S_2^n |Y_2^n) - H(S_1^n | W_0 W_2 S_2^n Y_2^n ) + H( S_1^n | W_0 S_2^n Y_1^n) -H( S_1^n | W_0 W_1 S_2^n Y_2^n) - n \epsilon_n \\
&\geq& H( S_1^n S_2^n |Y_2^n)- H(S_1^n| W_0 S_2^n Y_2^n ) + H( S_1^n | W_0 S_2^n Y_1^n)-H( S_1^n | W_0 W_1 S_2^n Y_2^n) - n\epsilon_n 
\end{IEEEeqnarray}
where $(a)$ stems from \eqref{Fano-inequality-1}. 
 
Let us define then: 
\begin{IEEEeqnarray}{rCl}
  A &\triangleq& H( S_1^n | W_0 S_2^n Y_1^n)  - H(S_1^n  | W_0 S_2^n Y_2^n ) \ ,  \\
  B &\triangleq& H( S_1^n | W_0 W_1 S_2^n Y_1^n)  \ . 
\end{IEEEeqnarray}
In the following, we aim at lower-bounding each of the two quantities $A$ and $B$. 

Let us start by writing that 
\begin{IEEEeqnarray}{rCl}
\IEEEyesnumber\IEEEyessubnumber* 
A &\triangleq& H( S_1^n | W_0 S_2^n Y_1^n)  - H(S_1^n  | W_0  S_2^n Y_2^n )  \\
&=& I( S_1^n; Y_2^n | W_0 S_2^n )  - I(S_1^n; Y_1^n | W_0  S_2^n ) \\
&=& \sum_{i=1}^n \bigl[  I(S_1^n; Y_{2,i} | W_0 Y_{2}^{i-1} S_2^n)  - I(S_1^n; Y_{1,i} | W_0  Y_{1,i+1}^n S_2^n) \bigr] \\ 
&\overset{(a)}{=}& \sum_{i=1}^n \bigl[  I(S_1^n Y_{1,i+1}^n ; Y_{2,i} | W_0 Y_{2}^{i-1}  S_2^n)   - I(S_1^n Y_{2}^{i-1}; Y_{1,i} | W_0 Y_{1,i+1}^n S_2^n) \bigr] \\ 
&\overset{(b)}{=}& \sum_{i=1}^n \bigl[  I(S_1^n; Y_{2,i} | W_0 Y_{2}^{i-1} Y_{1,i+1}^n S_2^n) - I(S_1^n; Y_{1,i} | W_0 Y_{2}^{i-1} Y_{1,i+1}^n S_2^n) \bigr] \\
&\overset{(c)}{=}&  \sum_{i=1}^n \bigl[  I(S_{1,i}; Y_{2,i} | W_0 Y_{2}^{i-1} Y_{1,i+1}^n S_2^n)  - I( S_{1,i}; Y_{1,i} | W_0 Y_{2}^{i-1} Y_{1,i+1}^n S_2^n) \bigr] \\
&=& \sum_{i=1}^n \bigl[  H(S_{1,i} | Y_{1,i} W_0 Y_{2}^{i-1} Y_{1,i+1}^n S_2^n)  - H( S_{1,i} |  Y_{2,i} W_0 Y_{2}^{i-1} Y_{1,i+1}^n S_2^n) \bigr] \\
&=& \sum_{i=1}^n \bigl[  H(S_{1,i} | Y_{1,i} S_{2,i} U_{0,i} ) - H(S_{1,i} |  Y_{2,i} S_{2,i} U_{0,i}) \bigr] 
\end{IEEEeqnarray} 

where $(a)$ follows using the following Csisz\'ar-K\"orner sum-identity
\begin{equation}
 \sum_{i=1}^n    I(Y_2^{i-1} ; Y_{1,i} | S_1^n W_0 Y_{1,i+1}^n S_2^n) = \sum_{i=1}^n I( Y_{1,i+1}^n; Y_{2,i} |S_1^n W_0  Y_2^{i-1} S_2^n), 
 \end{equation}
$(b)$ follows using the Csisz\'ar-K\"orner sum-identity given by 
\begin{IEEEeqnarray}{rCl}
  && \sum_{i=1}^n    I(Y_2^{i-1} ; Y_{1,i} | W_0  Y_{1,i+1}^n S_2^n) =  \sum_{i=1}^n I( Y_{1,i+1}^n; Y_{2,i} | W_0  Y_2^{i-1} S_2^n)  \ , 
 \end{IEEEeqnarray}
while $(c)$ is the consequence of the following sequence of Markov chains 
  \begin{IEEEeqnarray}{rCl}
  \IEEEyesnumber\IEEEyessubnumber* 
 (S_{1}^{i-1}, S_{1,i+1}^n,  S_{2}^{i-1}, S_{2,i+1}^n, Y_{1,i+1}^n, Y_2^{i-1} )  \mkv  (S_{1,i}, S_{2,i}) &\mkv&  Y_{j,i} \label{Markov-chain-iid-assumption_1}\\ 
\overset{(a)}{\Rightarrow} (S_{1}^{i-1}, S_{1,i+1}^n,  S_{2}^{i-1}, S_{2,i+1}^n, Y_{1,i+1}^n, Y_2^{i-1},  W_0 )  \mkv  (S_{1,i}, S_{2,i}) &\mkv&  Y_{j,i} \\
\Rightarrow (S_{1}^{i-1}, S_{1,i+1}^n)  \mkv (S_{2}^{i-1}, S_{2,i+1}^n, Y_{1,i+1}^n, Y_2^{i-1} , W_0, S_{1,i}, S_{2,i}) &\mkv& Y_{j,i} 
\end{IEEEeqnarray}
where \eqref{Markov-chain-iid-assumption_1} results from that the source sequences $(S_1^n, S_2^n, Y_1^n, Y_2^n)$ are memoryless, while $(a)$ is a consequence of that $W_0$ is a function of the pair of sequences $(S_1^n, S_2^n)$. 

To lower-bound the term B, note the following 
\begin{IEEEeqnarray}{rCl}
\IEEEyesnumber\IEEEyessubnumber* 
   B &\triangleq& H( S_1^n | W_0 W_1 S_2^n Y_1^n)  \\
   &=& \sum_{i=1}^n H( S_{1,i} | W_0 W_1 S_2^n Y_1^n S_1^{i-1}) \\
   &=& \sum_{i=1}^n H( S_{1,i} | S_{2,i} Y_{1,i} W_0 S_{2,<i>} Y_{1,i+1}^n S_1^{i-1} W_1 Y_1^{i-1}) \\
   &\overset{(a)}{=}& \sum_{i=1}^n H( S_{1,i} | S_{2,i} Y_{1,i} W_0 S_{2,<i>} Y_{1,i+1}^n S_1^{i-1} Y_2^{i-1} W_1 Y_1^{i-1}) \\
   &\leq& \sum_{i=1}^n H( S_{1,i} | S_{2,i} Y_{1,i} W_0 S_{2, <i>} Y_{1,i+1}^n Y_2^{i-1} W_1 Y_1^{i-1}) \label{bound_B}
\end{IEEEeqnarray}  
where $(a)$ is a consequence of the following sequence of Markov chains:
\begin{IEEEeqnarray}{rCl}
\IEEEyesnumber\IEEEyessubnumber* 
Y_{2}^{i-1} &\mkv& (S_{1}^{i-1} , S_{2}^{i-1} , Y_1^{i-1}) \mkv ( S_{1,i} ,S_{1,i+1}^n, S_{2,i}, S_{2,i+1}^n, Y_{1,i+1}^n) \label{Markov-chain-iid-assumption_2}\\
\overset{(a)}{\Rightarrow} Y_{2}^{i-1} &\mkv& ( S_{1}^{i-1} , S_{2}^{i-1}, Y_1^{i-1}) \mkv ( S_{1,i} , S_{1}^{i-1}, S_{2,i}, S_{2}^{i-1}, Y_{1,i+1}^n , W_0, W_1) \\
\Rightarrow Y_{2}^{i-1} &\mkv& ( S_{1}^{i-1} , S_{2}^{i-1},  Y_1^{i-1}, S_{2,i}, S_{2}^{i-1}, Y_{1,i+1}^n, W_0, W_1) \mkv S_{1,i} \ . 
\end{IEEEeqnarray}
where \eqref{Markov-chain-iid-assumption_2} results from that the source sequences $(S_1^n, S_2^n, Y_1^n , Y_2^n)$ are memoryless, while $(a)$ is a consequence of that $W_0$ and $W_1$ are each function of the pair of sequences $(S_1^n, S_2^n)$. 

Finally, letting $U_{1,i}  \triangleq (W_1, Y_1^{i-1})$ so that the choice of $(U_{0,i} , U_{1,i}) $ satisfy the condition: $ \hat{S}_{1,i} = g_i( Y_{1,i}, U_{0,i}, U_{1,i}, S_{2,i})  $, we write the resulting sum-rate constraint as 
\begin{IEEEeqnarray}{rCl} 
n (R_0 + R_1 + R_2)  &\geq& n H( S_1  S_2  |Y_2 )+ \sum_{i=1}^n \bigl[ H( S_{1,i} | S_{2,i} Y_{1,i}U_{0,i}) - H( S_{1,i} | S_{2,i} Y_{2,i} U_{0,i}) \bigr] \IEEEnonumber \\
&& \qquad \qquad - \sum_{i=1}^n H( S_{1,i} | S_{2,i} Y_{1,i} U_{0,i} U_{1,i}) -  n\epsilon_n 
\end{IEEEeqnarray}

Let us now prove that the following bound holds 
\begin{equation}
 R_0 + R_1  \geq  H(S_2 S_1 |Y_1 ) - H(S_1| U_0 U_1 Y_1 S_2) \ . 
\end{equation}

We have  
\begin{IEEEeqnarray}{rCl}
\IEEEyesnumber\IEEEyessubnumber* 
n (R_0 + R_1) &\geq& H(W_0) + H(W_1|W_0) \\ 
&=& H(W_0, W_1) \\
&\geq& H(W_0  W_1 |Y_1^n) \\
&\geq& I(W_0  W_1 ; S_1^nS_2^n|Y_1^n) \\
&=& H(S_1^nS_2^n|Y_1^n) - H( S_1^nS_2^n| W_0 W_1 Y_1^n)  \\
&\overset{(a)}{\geq}& H(S_1^nS_2^n|Y_1^n) - H( S_1^n| W_0 W_1 S_2^n Y_1^n)   - n\epsilon_n \\
&=& n H(S_1S_2| Y_1) - B  - n\epsilon_n \\
&\geq& n H(S_1S_2| Y_1) - \sum_{i=1}^n H( S_{1,i} | S_{2,i} Y_{1,i} U_{0,i} U_{1,i})  - n\epsilon_n  \ . 
\end{IEEEeqnarray}
where $(a)$ is a consequence of Fano's inequality in \eqref{Fano-inequality-1} and $(b) $ results from the lower bound on B in \eqref{bound_B}.

As for the third rate constraint
\begin{equation}
 R_0 + R_2 \geq H(S_1 S_2 |Y_2) - H( S_1 | U_0Y_2 S_2)  \ , 
\end{equation}
we write
\begin{IEEEeqnarray}{rCl}
\IEEEyesnumber\IEEEyessubnumber* 
n ( R_0 + R_2) &\geq& H(W_0 W_2)  \\ 
&\geq& H(W_0 W_2  |Y_2^n) \\
&\geq& I(W_0 W_2; S_1^nS_2^n|Y_2^n) \\  
&=& H(S_1^nS_2^n|Y_2^n ) - H(S_1^n S_2^n| W_0 W_2 Y_2^n) \\
&\overset{(a)}{\geq}& H(S_1^nS_2^n|Y_2^n )-  H( S_1^n |W_0 W_2 S_2^n Y_2^n)- n \epsilon_n  \\
&\geq& H(S_1^nS_2^n|Y_2^n ) - H( S_1^n |W_0 S_2^n Y_2^n) -  n \epsilon_n\\
&=& n H(S_1 S_2 | Y_2) - \sum_{i=1}^n H( S_{1,i} |  S_{2,i} Y_{2,i} W_0 S_{2,<i>} Y_{2,<i>} S_{1,i+1}^n) - n\epsilon_n \\   
&=& n  H(S_1 S_2 | Y_2) -  \sum_{i=1}^n H( S_{1,i} |  S_{2,i} Y_{2,i}W_0 S_{2,<i>} Y_{2,<i>} S_{1,i+1}^n Y_{1,i+1}^n  ) - n\epsilon_n  \\
&\geq& n H(S_1 S_2 | Y_2) -  \sum_{i=1}^n H( S_{1,i} |  S_{2,i} Y_{2,i} W_0 S_{2,<i>}  Y_{2}^{i-1} Y_{1,i+1}^n) - n\epsilon_n  \\
&=& n  H(S_1 S_2 | Y_2) - \sum_{i=1}^n H(S_{1,i} |  S_{2,i} Y_{2,i} U_{0,i}) - n\epsilon_n \ . 
\end{IEEEeqnarray}
where $(a)$ is a consequence of Fano's inequality in \eqref{Fano-inequality-2} and $(b)$ stems for the following sequence of Markov Chain 
\begin{IEEEeqnarray}{rCl}
\IEEEyesnumber\IEEEyessubnumber* 
Y_{1,i+1}^n  &\mkv& ( S_{2,i+1}^n ,S_{1,i+1}^n , Y_{1,i+1}^n ) \mkv ( S_{1,i} ,S_{1}^{i-1}, S_{2,i}, S_{2}^{i-1}, Y_1^{i-1}) \label{Markov-chain-iid-assumption_3}\\
\overset{(a)}{\Rightarrow} Y_{1,i+1}^n  &\mkv& ( S_{2,i+1}^n ,S_{1,i+1}^n , Y_{1,i+1}^n ) \mkv ( S_{1,i} ,S_{1}^{i-1}, S_{2,i}, S_{2}^{i-1}, Y_1^{i-1}, W_0, W_1) \\
\Rightarrow Y_{1,i+1}^n  &\mkv& ( S_{2,i+1}^n ,S_{1,i+1}^n , Y_{1,i+1}^n, S_{2,i}, S_{2}^{i-1}, Y_1^{i-1}, W_0, W_1) \mkv S_{1,i} \ . 
\end{IEEEeqnarray}
where \eqref{Markov-chain-iid-assumption_3} results from that the source sequences $(S_1^n, S_2^n, Y_1^n , Y_2^n)$ are memoryless, while $(a)$ is a consequence of that $W_0$ and $W_1$ are each function of the pair of sequences $(S_1^n, S_2^n)$.

\noindent Let $Q$ be an integer-valued random variable, ranging from $1$ to $n$, uniformly distributed over $[1:n]$ and independent of all other variables $(S_1,S_2, U_0, U_1,  Y_1 , Y_2) $. We have
\begin{IEEEeqnarray}{rCl}   
\IEEEyesnumber\IEEEyessubnumber* 
 R_0 + R_1 + R_2  &\geq  & H( S_1  S_2 |Y_2 )+ \dfrac{1}{n} \ \sum_{i=1}^n \bigl[ H( S_{1,i} | S_{2,i} Y_{1,i}U_{0,i}) - H( S_{1,i} | S_{2,i} Y_{2,i} U_{0,i}) \bigr] \IEEEnonumber \\
&& \qquad \qquad -  \dfrac{1}{n} \sum_{i=1}^n H( S_{1,i} | S_{2,i} Y_{1,i} U_{0,i} U_{1,i}) -  n\epsilon_n  \\
&=&H( S_1  S_2 |Y_2 )+  \ \sum_{i=1}^n \mathds{P}(Q= i)\bigl[ H( S_{1,Q} | S_{2,Q} Y_{1,Q} U_{0,Q} , Q= i) - H( S_{1,Q} | S_{2,Q} Y_{2,Q} U_{0,Q}, Q= i) \bigr] \IEEEnonumber \\
&& \qquad \qquad -   \sum_{i=1}^n \mathds{P}(Q= i) H( S_{1,Q} | S_{2,Q} Y_{1,Q} U_{0,Q} U_{1,Q}, Q= i ) -  n\epsilon_n  \\ 
&=&H( S_1  S_2 |Y_2 )+   H( S_{1,Q} | S_{2,Q} Y_{1,Q} U_{0,Q} Q) - H( S_{1,Q} | S_{2,Q} Y_{2,Q} U_{0,Q} Q)   \IEEEnonumber \\
&& \qquad \qquad -    H( S_{1,Q} | S_{2,Q} Y_{1,Q} U_{0,Q} U_{1,Q} Q) -  n\epsilon_n \ .  \\ 
&\overset{(a)}{=}& H( S_1  S_2 |Y_2 )+  H( S_{1} | S_{2} Y_{1} U_{0,Q} Q) - H( S_{1} | S_{2} Y_{2} U_{0,Q} Q)  \IEEEnonumber \\
&& \qquad \qquad -    H( S_{1} | S_{2} Y_{1} U_{0,Q} U_{1,Q} Q) -  n\epsilon_n \ .  
\end{IEEEeqnarray} 
where $(a)$ is a consequence of that all sources $(S_1^n, S_2^n, Y_1^n , Y_2^n)$ are memoryless.

\noindent Let us now define $U_1 \triangleq (Q, U_{1,Q})$ and $U_0 \triangleq (Q, U_{0,Q})$, we obtain 
  \begin{equation}
   R_0 + R_1 + R_2 \geq H( S_1  S_2 |Y_2 )+ H( S_{1} | S_{2} Y_{1} U_{0}) - H( S_{1} | S_{2} Y_{2} U_{0}) - H( S_{1} | S_{2} Y_{1} U_{0} U_{1}) \ . 
  \end{equation} 
The two other rate constraints can be written in a similar fashion, 
\begin{IEEEeqnarray}{rCl}   
\IEEEyesnumber\IEEEyessubnumber* 
R_0 + R_1  &\geq&  H(S_2 S_1 |Y_1 ) - H(S_1| U_0 U_1 Y_1 S_2) \\
R_0 + R_2 &\geq& H(S_1 S_2 |Y_2) - H( S_1 | U_0Y_2 S_2)  \ ; 
\end{IEEEeqnarray}
and this completes the proof of converse. \qed 

\subsection{Proof of Direct Part}

For the proof of achievability of Theorem~\ref{th-Gray-Wyner-side-information}, for convenience~\footnote{The readers who are well acquainted with coding for Heergard-Berger type and Gray-Wyner models, may find the region more appropriate (for intuitions) in its form of Proposition~\ref{proposition-proof-direct-part-theorem1}.} we first show that the rate-distortion region of the proposition that will follow is achievable. The achievability of the rate-distortion region of Theorem~\ref{th-Gray-Wyner-side-information} follows by choosing the random variable $V_0$ of the proposition as $V_0=(U_0,S_2)$.
 
\begin{proposition}\label{proposition-proof-direct-part-theorem1}
An inner bound on the rate-distortion region of the Gray-Wyner model with side information and degraded reconstruction sets of Figure \ref{fig-Gray-Wyner-model-with-side-info-degraded-reconstructions} is given by the set of all rate-distortion quadruples $(R, R_1, R_2, D_1)$ that satisfy  
\begin{subequations}
\begin{align} 
R_0 + R_1 &\geq I(V_0 U_1; S_1 S_2 | Y_1) \\
R_0 + R_2 &\geq I(V_0 ; S_1 S_2 | Y_2) \\
R_0 + R_1 + R_2 &\geq \max \: \{ I(V_0 ; S_1 S_2 |  Y_1), I(V_0 ; S_1 S_2 |  Y_2) \} +  I(U_1 ; S_1 S_2| V_0 Y_1)
\end{align}
\end{subequations}
for some choice of the random variables $(V_0,U_1)$ such that $(V_0,U_1) \mkv (S_1,S_2) \mkv (Y_1, Y_2)$ and there exist functions $g_1$, $g_{2,1}$, and $g_{2,2}$ such that: 
\begin{subequations}
\begin{align} 
 \hat{S}_1 &= g_1(V_0, U_1, Y_1) \\
 S_2 &= g_{2,1}( V_0, U_1, Y_1) \\
 S_2 &= g_{2,2}( V_0, Y_2)  \ , 
\end{align}
\label{reconstruction-constraint1-proposition-proof-direct-part-theorem1}
\end{subequations}
and 
\begin{equation}
\mathds{E} d_1(S_1; \hat{S}_1)\leq D_1.
\label{reconstruction-constraint2-proposition-proof-direct-part-theorem1}
\end{equation} 
\end{proposition}

\noindent \textbf{Proof of Proposition~\ref{proposition-proof-direct-part-theorem1}:} We now describe a coding scheme that achieves the rate-distortion region of Proposition~\ref{proposition-proof-direct-part-theorem1}. The scheme is very similar to one that is developed by Shayevitz and Wigger~\cite[Theorem 2]{Shayevitz2013} for a Gray-Wyner model with side information. In particular, similar to ~\cite[Theorem 2]{Shayevitz2013} it uses a double-binning technique for the common codebook, one that is relevant for Receiver $1$ and one that is relevant for Receiver $2$. Note, however, that, formally, the result of Proposition~\ref{proposition-proof-direct-part-theorem1} cannot be obtained by readily applying ~\cite[Theorem 2]{Shayevitz2013} as is; and one needs to extend the result of ~\cite[Theorem 2]{Shayevitz2013} in a manner that accounts for that the source component $S^n_2$ is to be recovered losslessly by both decoders. This can be obtained by extending the distortion measure of ~\cite[Theorem 2]{Shayevitz2013} to one that is vector-valued, i.e., $d\left((s_1,s_2),(\hat{s}_1,\hat{s}_2)\right)=\left(d_1(s_1,\hat{s}_1),d_H(s_2,\hat{s}_2)\right)$, where $d_H(\cdot,\cdot)$ denotes the Hamming distance. For reasons of completeness, we provide here an outline proof of  Proposition~\ref{proposition-proof-direct-part-theorem1}.

Our scheme has the following parameters: a conditional joint measure $P_{V_0U_1|S_1S_2}$ that satisfies \eqref{reconstruction-constraint1-proposition-proof-direct-part-theorem1} and \eqref{reconstruction-constraint2-proposition-proof-direct-part-theorem1}, and non-negative communication rates $T_0$, $T_1$, $T_{0,0}$, $T_{0,p}$, $T_{1,0}$, $T_{1,1}$, $\tilde{R}_{0,0}$, $\tilde{R}_{0,1}$, $\tilde{R}_{0,2}$, $\tilde{R}_{1,0}$ and $\tilde{R}_{1,1}$ such that

\begin{subequations}
\begin{align} 
  T_0 &= T_{0,0} + T_{0,p} \quad ,  \quad 0  \leq  \tilde{R}_{0,0} \leq T_{0,0} \quad ,  \quad  0  \leq \tilde{R}_{0,1} \leq T_{0,p}  \quad ,  \quad  0  \leq  \tilde{R}_{0,2} \leq T_{0,p}\\
  T_1 &= T_{1,0} + T_{1,1} \quad ,  \quad  0  \leq  \tilde{R}_{1,0} \leq T_{1,0}\quad ,  \quad  0  \leq \tilde{R}_{1,1} \leq T_{1,1}.
\end{align}
\label{nuisance-variables-relations}
\end{subequations}

\subsubsection*{Codebook Generation} 
\begin{itemize}
\item[1)] Randomly and independently generate $2^{n T_0}$ length-$n$ codewords $v_0^n(k_0)$ indexed with the pair of indices $k_0=(k_{0,0},k_{0,p})$, where $k_{0,0} \in [1:2^{nT_{0,0}}]$ and $k_{0,p} \in [1:2^{nT_{0,p}}]$. Each codeword $v_0^n(k_0)$ has i.i.d entries drawn according to  $\displaystyle\prod_{i=1}^n P_{V_0}(v_{0,i}(k_0))$. The codewords $\{v_0^n(k_0)\}$ are partitioned into superbins whose indices will be relevant for both receivers; and each superbin is partioned int two different ways, each into subbins whose indices will be relevant for a distinct receiver (i.e., double-binning). This is obtained by partitioning the indices $\{(k_{0,0}, k_{0,p})\}$ as follows.  We partition the $2^{nT_{0,0}}$ indices $\{k_{0,0}\}$ into $2^{n \tilde{R}_{0,0}}$ bins by randomly and independently assigning each index $k_{0,0}$ to an index $\tilde{w}_{0,0}(k_{0,0})$ according to a uniform pmf over $[1: 2^{n \tilde{R}_{0,0}}]$. We refer to each subset of indices $\{k_{0,0}\}$ with the same index $\tilde{w}_{0,0}$ as a bin $\mc B_{00}(\tilde{w}_{0,0})$, $\tilde{w}_{0,0} \in [1: 2^{n \tilde{R}_{0,0}}]$. Also, we make two distinct partitions of the $2^{nT_{0,p}}$ indices $\{k_{0,p}\}$, each relevant for a distinct receiver. In the first partition, which is relevant for Receiver $1$, the indices $\{k_{0,p}\}$ are assigned randomly and independently each to an index $\tilde{w}_{0,1}(k_{0,p})$ according to a uniform pmf over $[1: 2^{n \tilde{R}_{0,1}}]$. We refer to each subset of indices $\{k_{0,p}\}$ with the same index $\tilde{w}_{0,1}$ as a bin $\mc B_{01}(\tilde{w}_{0,1})$, $\tilde{w}_{0,1} \in [1: 2^{n \tilde{R}_{0,1}}]$. Similarly, in the second partition, which is relevant for Receiver $2$, the indices $\{k_{0,p}\}$ are assigned randomly and independently each to an index $\tilde{w}_{0,2}(k_{0,p})$ according to a uniform pmf over $[1: 2^{n \tilde{R}_{0,2}}]$; and refer to each subset of indices $\{k_{0,p}\}$ with the same index $\tilde{w}_{0,2}$ as a bin $\mc B_{02}(\tilde{w}_{0,2})$, $\tilde{w}_{0,2} \in [1: 2^{n \tilde{R}_{0,2}}]$.
 
\item[2)] For each $k_0 \in [1: 2^{nT_0}]$, randomly and independently generate $2^{n T_1}$ length-$n$ codewords $u_1^n(k_1, k_0)$ indexed with the pair of indices $k_1=(k_{1,0},k_{1,1})$, where $k_{1,0} \in [1:2^{nT_{1,0}}]$ and $k_{1,1} \in [1:2^{nT_{1,1}}]$. Each codeword $u_1^n(k_1, k_0)$ is with i.i.d elements drawn according to  $\displaystyle\prod_{i=1}^n P_{U_1|V_0}(u_{1,i}(k_1, k_0) |v_{0,i}(k_0))$.  We partition the $2^{nT_{1,0}}$ indices $\{k_{1,0}\}$ into $2^{n \tilde{R}_{1,0}}$ bins by randomly and independently assigning each index $k_{1,0}$ to an index $\tilde{w}_{1,0}(k_{1,0})$ according to a uniform pmf over $[1: 2^{n \tilde{R}_{1,0}}]$. We refer to each subset of indices $\{k_{1,0}\}$ with the same index $\tilde{w}_{1,0}$ as a bin $\mc B_{10}(\tilde{w}_{1,0})$, $\tilde{w}_{1,0} \in [1: 2^{n \tilde{R}_{1,0}}]$. Similarly, we partition the $2^{nT_{1,1}}$ indices $\{k_{1,1}\}$ into $2^{n \tilde{R}_{1,1}}$ bins by randomly and independently assigning each index $k_{1,1}$ to an index $\tilde{w}_{1,1}(k_{1,1})$ according to a uniform pmf over $[1: 2^{n \tilde{R}_{1,1}}]$; and refer to each subset of indices $\{k_{1,1}\}$ with the same index $\tilde{w}_{1,1}$ as a bin $\mc B_{11}(\tilde{w}_{1,1})$, $\tilde{w}_{1,1} \in [1: 2^{n \tilde{R}_{1,1}}]$.

\item[3)] Reveal all codebooks and its partitions to the encoder, the codebook of $\{v_0^n(k_0)\}$ and its partitions to both receivers, and the codebook of $\{u_1^n(k_1, k_0)\}$ and its partitions to only Receiver $1$.

\end{itemize}
  
\subsubsection*{Encoding}
Upon observing the source pair $(S^n_1,S^n_2)=(s_1^n, s_2^n)$, the encoder finds an index $k_0=(k_{0,0},k_{0,p})$ such that the codeword  $v_0^n(k_0)$ is jointly typical with $(s^n_1,s^n_2)$, i.e.,
\begin{equation}
  \left( s_1^n, s_2^n,v_0^n(k_0)  \right) \in \mathcal{T}^{(n)}_{[S_1S_2V_0]} \ . 
\end{equation}
By the covering lemma~\cite[Chapter 3]{GK11}, the encoding in this step is successful as long as $n$ is large and
\begin{equation}
T_0  \geq I(V_0 ; S_1S_2).
\label{rate-constraint-encoding-step1-proof-proposition}
\end{equation}
Next, it finds an index $k_1=(k_{1,0},k_{1,1})$ such that the codeword $u_1^n(k_1, k_0)$ is jointly typical with the triple $(s_1^n, s_2^n, v_0^n(k_0))$, i.e.,  
\begin{equation}
  \left( s_1^n, s_2^n, v_0^n(k_0), u_1^n(k_1, k_0) \right) \in \mathcal{T}^{(n)}_{[S_1S_2V_0U_1]}.  
\end{equation}
Again, by the covering lemma~\cite[Chapter 3]{GK11}, the encoding in this step is successful as long as $n$ is large and
\begin{equation}
T_1  \geq I(U_1 ; S_1S_2|V_0).
\label{rate-constraint-encoding-step2-proof-proposition}
\end{equation}

\noindent Let $\tilde{w}_{0,0}$, $\tilde{w}_{0,1}$ and $\tilde{w}_{0,2}$ be the bin indices such that $k_{0,0} \in \mc B_{00}(\tilde{w}_{0,0})$, $k_{0,p} \in \mc B_{01}(\tilde{w}_{0,1})$ and $k_{0,p} \in \mc B_{02}(\tilde{w}_{0,2})$. Also, let $\tilde{w}_{1,0}$ and $\tilde{w}_{1,1}$ be the bin indices such that $k_{1,0} \in \mc B_{10}(\tilde{w}_{1,0})$ and $k_{1,1} \in \mc B_{11}(\tilde{w}_{1,1})$. The encoder then sends the product message $W_0 = (\tilde{w}_{0,0},\tilde{w}_{1,0})$ over the error-free rate-limited common link of capacity $R_0$. Also, it sends the product message $W_1= (\tilde{w}_{0,1},\tilde{w}_{1,1})$ over the error-free rate-limited individual link to Receiver $1$ of capacity $R_1$, and the message $W_2=\tilde{w}_{0,2}$ over the error-free rate-limited individual link to Receiver $2$ of capacity $R_2$.

\subsubsection*{Decoding}

\noindent Receiver $1$ gets the messages $(W_0,W_1)=(\tilde{w}_{0,0},\tilde{w}_{1,0},\tilde{w}_{0,1},\tilde{w}_{1,1})$. It seeks a codeword $v_0^n(k_0)$ and a codeword $u^n_1(k_1, k_0)$, with the indices $k_0=(k_{0,0},k_{0,p})$ and $k_1=(k_{1,0}, k_{1,1})$ satisfying $k_{0,0} \in \mc B_{00}(\tilde{w}_{0,0})$, $k_{0,p} \in \mc B_{01}(\tilde{w}_{0,1})$, $k_{1,0} \in \mc B_{10}(\tilde{w}_{1,0})$ and $k_{1,1} \in \mc B_{11}(\tilde{w}_{1,1})$, and such that 
\begin{equation}
  \left(  v_0^n(k_0),  u^n_1(k_1, k_0) , y_1^n  \right) \in \mathcal{T}^{(n)}_{[V_0 U_1 Y_1]} \ . 
\end{equation}

\noindent By the multivariate packing lemma~\cite[Chapter 12]{GK11}, the error in this decoding step at Receiver $1$ vanishes exponentially as long as $n$ is large and   
\begin{subequations}
\begin{align}
  T_{0,0} - \tilde{R}_{0,0} + T_{0,p} - \tilde{R}_{0,1}  &\leq I(V_0 ; Y_1) \\
  T_{1,0} - \tilde{R}_{1,0} + T_{1,1} - \tilde{R}_{1,1}  &\leq I(U_1 ; Y_1 |V_0) \ . 
\end{align}
\label{rate-constraint-decoding-receiver1-proof-proposition}
\end{subequations}

\noindent Receiver $1$ then sets its reproduced codewords $\hat{s}^n_{2,1}$ and $\hat{s}^n_1$ respectively as
\begin{subequations}
\begin{align}
\hat{s}^n_{2,1} &= g_{2,1}\left(v_0^n(k_0) , u^n_1(k_1, k_0) , y_1^n \right)\\
\hat{s}_1^n &= g_1 \left(v_0^n(k_0),  u^n_1(k_1, k_0) , y_1^n  \right)\ .
\end{align}
\end{subequations}

\noindent Similary, Receiver $2$ gets the message $(W_0,W_2)=(\tilde{w}_{0,0},\tilde{w}_{1,0},\tilde{w}_{0,2})$. It seeks a codeword $v_0^n(k_0)$, with $k_0=(k_{0,0},k_{0,p})$ satisfying $k_{0,0} \in \mc B_{00}(\tilde{w}_{0,0})$ and $k_{0,p} \in \mc B_{02}(\tilde{w}_{0,2})$, and such that 
\begin{equation}
  \left( v_0^n(k_0), y_1^n  \right) \in \mathcal{T}^{(n)}_{[V_0 Y_2]} \ . 
\end{equation}

\noindent Again, using the multivariate packing lemma~\cite[Chapter 12]{GK11}, the error in this decoding step at Receiver $2$ vanishes exponentially as long as $n$ is large and   
\begin{equation} 
  T_{0,0} - \tilde{R}_{0,0} + T_{0,p} - \tilde{R}_{0,2}  \leq I(V_0 ; Y_2).
	\label{rate-constraint-decoding-receiver2-proof-proposition}
\end{equation}
\noindent Receiver $2$ then ets its reconstructed codeword $\hat{s}^n_{2,1}$ as
\begin{equation}
\hat{s}^n_{2,2} = g_{2,2} \left(v_0^n(k_0), y_2^n  \right)\ . 
\end{equation}

\noindent Summarizing, combining \eqref{rate-constraint-encoding-step1-proof-proposition}, \eqref{rate-constraint-encoding-step2-proof-proposition}, \eqref{rate-constraint-decoding-receiver1-proof-proposition} and \eqref{rate-constraint-decoding-receiver2-proof-proposition}, the communication rates $T_0$, $T_1$, $T_{0,0}$, $T_{0,p}$, $T_{1,0}$, $T_{1,1}$, $\tilde{R}_{0,0}$, $\tilde{R}_{0,1}$, $\tilde{R}_{0,2}$, $\tilde{R}_{1,0}$ and $\tilde{R}_{1,1}$ satisfy the following inequalities
\begin{subequations}
\begin{align} 
 	T_0  &\geq I(V_0 ; S_1S_2) \\
  T_1  &\geq I(U_1 ; S_1S_2|V_0) \\
 	T_{0,0} - \tilde{R}_{0,0} + T_{0,p} - \tilde{R}_{0,1}  &\leq I(V_0 ; Y_1) \\
  	T_{0,0} - \tilde{R}_{0,0} + T_{0,p} - \tilde{R}_{0,2}  &\leq I(V_0 ; Y_2) \\
  	T_{1,0} - \tilde{R}_{1,0} + T_{1,1} - \tilde{R}_{1,1}  &\leq I(U_1  ; Y_1 |V_0).
\end{align}
\label{encoding-decoding-rate-constraints-proof-proposition}
\end{subequations}
\noindent Choosing $\tilde{R}_{0,0}$, $\tilde{R}_{1,1}$, $\tilde{R}_{0,2}$, $\tilde{R}_{1,0}$ and $\tilde{R}_{1,1}$ to also satisfy the  rate relations
\begin{subequations}
\begin{align}
    R_0 &= \tilde{R}_{0,0} + \tilde{R}_{1,0}  \\
    R_1 &= \tilde{R}_{0,1} + \tilde{R}_{1,1} \\
    R_2 &= \tilde{R}_{0,2}.
\end{align}
\label{binning-rates-relations}
\end{subequations}
and, finally, using Fourier-Motzkin elimination (FME) to successively project out the nuisance variables $T_{0,0}$, $T_{0,p}$, $T_{1,0}$, $T_{1,1}$, $T_0$, $T_1$, and then $\tilde{R}_{0,0}$, $\tilde{R}_{0,1}$, $\tilde{R}_{0,2}$, $\tilde{R}_{1,0}$ and $\tilde{R}_{1,1}$ from the set of relations formed by~\eqref{nuisance-variables-relations}, \eqref{encoding-decoding-rate-constraints-proof-proposition} and \eqref{binning-rates-relations}, we get the region of Proposition~\ref{proposition-proof-direct-part-theorem1}.

\noindent This completes the proof of the proposition; and so that of the direct part of Theorem \ref{th-Gray-Wyner-side-information}. 


\bibliographystyle{IEEEtran}
\bibliography{BiblioSourceCoding}

\end{document}